\input amssym
\input harvmac
\input texdraw

\noblackbox

\def\mn{\the\secno.\the\subsecno}
\baselineskip=16pt plus 2pt minus 1pt
\parskip=2pt plus 16pt minus 1pt

%
\def\lfig#1{
\let\labelflag=#1%
\def\numb@rone{#1}%
\ifx\labelflag\UnDeFiNeD%
{\xdef#1{\the\figno}%
\writedef{#1\leftbracket{\the\figno}}%
\global\advance\figno by1%
}\fi{\hyperref{}{figure}{{\numb@rone}}{Fig.~{\numb@rone}}}}
\def\figboxinsert#1#2#3{%
\let\flag=#1
\ifx\flag\UnDeFiNeD
  {\xdef#1{\the\figno}
   \writedef{#1\leftbracket{\the\figno}}
   \global\advance\figno by1}
\fi
\vbox{\bigskip \centerline{#3} \smallskip
\leftskip 4pc \rightskip 4pc
\noindent\ninepoint\sl \baselineskip=11pt
{\bf{\hyperdef\hypernoname{figure}{{#1}}{Fig.~{#1}}}.~}#2
\smallskip}
\bigskip}

\def\DrawDiag#1{
{\btexdraw
\drawdim pt \linewd 0.75 \lpatt ()
\arrowheadtype t:F \arrowheadsize l:8 w:4
\textref h:C v:C
#1
\etexdraw} }

\newif\ifhypertex
\ifx\hyperdef\UnDeFiNeD
\hypertexfalse
\message{[HYPERTEX MODE OFF]}
\def\hyperref#1#2#3#4{#4}
\def\hyperdef#1#2#3#4{#4}
\def\hypernoname{}
\def\e@tf@ur#1{}

\else
\hypertextrue
\message{[HYPERTEX MODE ON]}

\fi

\def\IC{{\Bbb C}}

\def\IP{{\Bbb P}}

\def\IR{{\Bbb R}}

\def\IZ{{\Bbb Z}}


\def\CK {{\cal K}}
\def\CL {{\cal L}}

\def\CO {{\cal O}}

\def\CU {{\cal U}}
\def\CV {{\cal V}}

\def\bar{\overline}

\def\yyy{} 

\def\dd{{\rm d}} 
\def\ac{\vartheta_1}  
\def\bc{\vartheta_2}  
\def\UU{\alpha}    
\def\VV{\beta}       
\def\XX{\tilde\alpha} 
\def\YY{\tilde\beta}  


\lref\WittenJones{
E.~Witten, ``Quantum Field Theory And The Jones Polynomial,''
Commun.\ Math.\ Phys.\  {\bf 121}, 351 (1989).}

\lref\Wittencsstring{E.~Witten,
``Chern-Simons gauge theory as a string theory,''
Prog.\ Math.\  {\bf 133} (1995) 637, hep-th/9207094.}

\lref\GopakumarVafa{R.~Gopakumar, C.~Vafa,
``On the Gauge Theory/Geometry Correspondence,''
Adv.\ Theor.\ Math.\ Phys. {\bf 3} (1999) 1415.}

\lref\OV{H.~Ooguri, C.~Vafa, ``Knot Invariants and Topological Strings,''
Nucl.Phys. {\bf B577} (2000) 419.}

\lref\LMV{J.~M.~F.~Labastida, M.~Mari\~no and C.~Vafa,
  ``Knots, links and branes at large N,''
  JHEP {\bf 0011}, 007 (2000)
  [hep-th/0010102].}

\lref\BEM{A.~Brini, B.~Eynard and M.~Mari\~no,
  ``Torus knots and mirror symmetry,''
  Annales Henri Poincare {\bf 13}, 1873 (2012)
  [arXiv:1105.2012 [hep-th]].}

\lref\AKV{M.~Aganagic, A.~Klemm and C.~Vafa,
  ``Disk instantons, mirror symmetry and the duality web,''
  Z.\ Naturforsch.\ A {\bf 57}, 1 (2002)
  [hep-th/0105045].}

\lref\EO{B.~Eynard and N.~Orantin,
  ``Invariants of algebraic curves and topological expansion,''
  math-ph/0702045.}

\lref\HV{K.~Hori and C.~Vafa,
  ``Mirror symmetry,''
  hep-th/0002222.}

\lref\AV{M.~Aganagic and C.~Vafa,
  ``Mirror symmetry, D-branes and counting holomorphic discs,''
  hep-th/0012041.}

\lref\AKMV{M.~Aganagic, A.~Klemm, M.~Mari\~no and C.~Vafa,
  ``The Topological vertex,''
  Commun.\ Math.\ Phys.\  {\bf 254}, 425 (2005)
  [hep-th/0305132].}

\lref\ADKMV{M.~Aganagic, R.~Dijkgraaf, A.~Klemm, M.~Mari\~no and C.~Vafa,
  ``Topological strings and integrable hierarchies,''
  Commun.\ Math.\ Phys.\  {\bf 261}, 451 (2006)
  [hep-th/0312085].}

\lref\LM{J.~M.~F.~Labastida and M.~Mari\~no,
  ``Polynomial invariants for torus knots and topological strings,''
  Commun.\ Math.\ Phys.\  {\bf 217}, 423 (2001)
  [hep-th/0004196].}

\lref\MV{M.~Mari\~no and C.~Vafa,
  ``Framed knots at large N,''
  hep-th/0108064.}

\lref\ORV{A.~Okounkov, N.~Reshetikhin and C.~Vafa,
  ``Quantum Calabi-Yau and classical crystals,''
  Progr.\ Math.\  {\bf 244}, 597 (2006)
  [hep-th/0309208].}

\lref\GIKV{S.~Gukov, A.~Iqbal, C.~Kozcaz and C.~Vafa,
  ``Link Homologies and the Refined Topological Vertex,''
  Commun.\ Math.\ Phys.\  {\bf 298}, 757 (2010)
  [arXiv:0705.1368 [hep-th]].}

\lref\MacD{I.~G.~Macdonald,
   ``Symmetric Functions and Hall Polynomials,''
   Oxford University Press, 1995.}

\lref\KPW{D.~Krefl, S.~Pasquetti and J.~Walcher,
  ``The Real Topological Vertex at Work,''
  Nucl.\ Phys.\ B {\bf 833}, 153 (2010)
  [arXiv:0909.1324 [hep-th]].}

\lref\GMM{E.~Guadagnini, M.~Martellini and M.~Mintchev,
  ``Wilson Lines in Chern-Simons Theory and Link Invariants,''
  Nucl.\ Phys.\ B {\bf 330}, 575 (1990).}

\lref\BFMone{V.~Bouchard, B.~Florea and M.~Mari\~no,
  ``Counting higher genus curves with crosscaps in Calabi-Yau orientifolds,''
  JHEP {\bf 0412}, 035 (2004)
  [hep-th/0405083].}

\lref\BFMtwo{V.~Bouchard, B.~Florea and M.~Mari\~no,
  ``Topological open string amplitudes on orientifolds,''
  JHEP {\bf 0502}, 002 (2005)
  [hep-th/0411227].}

\lref\AVtwo{M.~Aganagic and C.~Vafa,
  ``Large N Duality, Mirror Symmetry, and a Q-deformed A-polynomial for Knots,''
  arXiv:1204.4709 [hep-th].}

\lref\Stev{S.~Stevan,
  ``Chern-Simons Invariants of Torus Links,''
  Annales Henri Poincare {\bf 11}, 1201 (2010)
  [arXiv:1003.2861 [hep-th]].}

\lref\RJ{M.~Rosso and V.~F.~R.~Jones,
   ``On the invariants of torus knots derived from quantum groups,''
   J. Knot Theory Ramifications {\bf 2} (1993).}

\lref\GV{R.~Gopakumar and C.~Vafa,
  ``On the gauge theory / geometry correspondence,''
  Adv.\ Theor.\ Math.\ Phys.\  {\bf 3}, 1415 (1999)
  [hep-th/9811131].}

\lref\DSV{D.~E.~Diaconescu, V.~Shende and C.~Vafa,
  ``Large N duality, Lagrangian cycles, and algebraic knots,''
  arXiv:1111.6533 [hep-th].}

\lref\AKMVtwo{M.~Aganagic, A.~Klemm, M.~Mari\~no and C.~Vafa,
  ``Matrix model as a mirror of Chern-Simons theory,''
  JHEP {\bf 0402}, 010 (2004)
  [hep-th/0211098].}

\lref\BCY{J.~Bryan, C.~Cadman and B.~Young,
  ``The Orbifold Topological Vertex,''
  arXiv:1008.4205 [math.AG].}

\lref\KPW{D.~Krefl, S.~Pasquetti and J.~Walcher,
  ``The Real Topological Vertex at Work,''
  Nucl.\ Phys.\ B {\bf 833}, 153 (2010)
  [arXiv:0909.1324 [hep-th]].}

\lref\IKV{A.~Iqbal, C.~Kozcaz and C.~Vafa,
  ``The Refined topological vertex,''
  JHEP {\bf 0910}, 069 (2009)
  [hep-th/0701156].}

\lref\IK{ A.~Iqbal and C.~Kozcaz,
  ``Refined Hopf Link Revisited,''
  JHEP {\bf 1204}, 046 (2012)
  [arXiv:1111.0525 [hep-th]].}

\lref\FGS{H.~Fuji, S.~Gukov, P.~Sulkowski and H.~Awata,
  ``Volume Conjecture: Refined and Categorified,''
  arXiv:1203.2182 [hep-th].}

\lref\AS{ M.~Aganagic and S.~Shakirov,
  ``Knot Homology from Refined Chern-Simons Theory,''
  arXiv:1105.5117 [hep-th].}

\lref\GAF{P.~Dunin-Barkowski, A.~Mironov, A.~Morozov, A.~Sleptsov and A.~Smirnov,
  ``Superpolynomials for toric knots from evolution induced by cut-and-join operators,''
  arXiv:1106.4305 [hep-th]; S.~Shakirov,
  ``$\beta$-Deformation and Superpolynomials of (n,m) Torus Knots,''
  arXiv:1111.7035 [math-ph]; A.~Mironov, A.~Morozov and S.~.Shakirov,
  ``Torus HOMFLY as the Hall-Littlewood Polynomials,''
  arXiv:1203.0667 [hep-th].}

\lref\DFone{D.~-E.~Diaconescu and B.~Florea,
  ``Large N duality for compact Calabi-Yau threefolds,''
  Adv.\ Theor.\ Math.\ Phys.\  {\bf 9}, 31 (2005)
  [hep-th/0302076].}

\lref\DFtwo{D.~-E.~Diaconescu and B.~Florea,
  ``Localization and gluing of topological amplitudes,''
  Commun.\ Math.\ Phys.\  {\bf 257}, 119 (2005)
  [hep-th/0309143].}

\lref\ChCh{L.~Chen and Q.~Chen,
  ``Orthogonal Quantum Group Invariants of Links,''
  arXiv:1007.1656 [math.QA].}

\lref\LLR{J.~M.~F.~Labastida, P.~M.~Llatas and A.~V.~Ramallo,
  ``Knot operators in Chern-Simons gauge theory,''
  Nucl.\ Phys.\ B {\bf 348}, 651 (1991).}

\lref\WittCSA{E.~Witten,
  ``Chern-Simons gauge theory as a string theory,''
  Prog.\ Math.\  {\bf 133}, 637 (1995)
  [hep-th/9207094].}

\lref\GV{R.~Gopakumar and C.~Vafa,
  ``On the gauge theory / geometry correspondence,''
  Adv.\ Theor.\ Math.\ Phys.\  {\bf 3}, 1415 (1999)
  [hep-th/9811131].}

\lref\MRev{M.~Mari\~no,
  ``Chern-Simons Theory, the 1/N Expansion, and String Theory,''
  arXiv:1001.2542 [hep-th].}

\lref\GVM{R.~Gopakumar and C.~Vafa,
  ``M theory and topological strings. 2.,''
  hep-th/9812127; R.~Gopakumar and C.~Vafa,
  ``M theory and topological strings. 1.,''
  hep-th/9809187.}

\lref\Kosh{S.~Koshkin,
  ``Conormal bundles to knots and the Gopakumar-Vafa conjecture,''
  math/0503248.}

\lref\Taub{C.~H.~Taubes,
  ``Lagrangians for the Gopakumar-Vafa conjecture,''
  Adv.\ Theor.\ Math.\ Phys.\  {\bf 5}, 139 (2001)
  [math/0201219 [math-dg]].}

\lref\KL{ S.~H.~Katz and C.~-C.~M.~Liu,
  ``Enumerative geometry of stable maps with Lagrangian boundary conditions and multiple covers of the disc,''
  Adv.\ Theor.\ Math.\ Phys.\  {\bf 5}, 1 (2002)
  [math/0103074 [math-ag]].}

\lref\BKMP{V.~Bouchard, A.~Klemm, M.~Mari\~no and S.~Pasquetti,
  ``Remodeling the B-model,''
  Commun.\ Math.\ Phys.\  {\bf 287}, 117 (2009)
  [arXiv:0709.1453 [hep-th]].}

\lref\LS{
  J.~Li and Y.~S.~Song,
  ``Open string instantons and relative stable morphisms,''
  Adv.\ Theor.\ Math.\ Phys.\  {\bf 5}, 67 (2002).
  [hep-th/0103100].
}

\lref\DV{R.~Dijkgraaf and C.~Vafa,
  ``Matrix models, topological strings, and supersymmetric gauge theories,''
  Nucl.\ Phys.\ B {\bf 644}, 3 (2002)
  [hep-th/0206255].}

\lref\MM{M.~Mari\~no,
  ``Open string amplitudes and large order behavior in topological string theory,''
  JHEP {\bf 0803}, 060 (2008)
  [hep-th/0612127].}

\lref\WProg{H.~Jockers, A.~Klemm and M.~Soroush, work in progress.}

\lref\Maul{D.~Maulik,
    ``Stable pairs and the HOMFLY polynomial,''
    arXiv:1210.6323 [math.AG].}

\lref\CKK{J.~Choi, S.~Katz and A.~Klemm,
  ``The refined BPS index from stable pair invariants,''
  arXiv:1210.4403 [hep-th].}

\lref\FujiNX{
H.~Fuji, S.~Gukov and P.~Su{\l}kowski,
``Super-A-polynomial for knots and BPS states,''
Nucl.\ Phys.\ B {\bf 867}, 506 (2013).
[arXiv:1205.1515 [hep-th]].
}

\lref\Ng{
L.~Ng,
``Framed knot contact homology,''
Duke\ Math.\ J. {\bf 141}, 365-406 (2008).
[math/0407071];
T.~Ekholm, J.~Etnyre, L.~Ng and M.~Sullivan,
``Filtrations on the knot contact homology of transverse knots,''
arXiv:1010.0450 [math.SG];
L.~Ng,
``Combinatorial knot contact homology and transverse knots,''
Adv.\ Math. {\bf 227}, No.~6, 2189-2219 (2011).
[arXiv:1010.0451 [math.SG]];
}

\lref\NgJX{
L.~Ng,
``A topological introduction to knot contact homology,''
[arXiv:1210.4803 [math.GT]].
}

\lref\Br{A.~Brini,
  ``Open topological strings and integrable hierarchies: Remodeling the A-model,''
  Commun.\ Math.\ Phys.\  {\bf 312}, 735 (2012)
  [arXiv:1102.0281 [hep-th]].}

\lref\BC{A.~Brini and R.~Cavalieri,
  ``Open orbifold Gromov-Witten invariants of $[C^3/Z_n]$: Localization and mirror symmetry,''
  arXiv:1007.0934 [math.AG].}

\lref\BCprog{A.~Brini and T.~Coates,
   ``The remodeled A-model and torus knots at large N,'' work in progress.}


\Title{\vbox{\baselineskip12pt\hbox{Bonn-TH-12-26}}}
{\vbox{
\centerline{Torus Knots and the Topological Vertex}
}}
\centerline{Hans Jockers, Albrecht Klemm, and Masoud Soroush}
\medskip
\vskip 8pt
\centerline{\it Bethe Center for Theoretical Physics, Physikalisches Institut}
\centerline{\it der Universit\"{a}t Bonn, Nussallee 12, Bonn D-53315, Germany}
\vskip 30pt
{\bf \centerline{Abstract}}

We propose a class of toric Lagrangian A-branes on the resolved conifold that is suitable to describe torus knots on $S^3$. The key role is played by the $SL(2,\IZ)$ transformation, which generates a general torus knot from the unknot. Applying the topological vertex to the proposed A-branes, we rederive the colored HOMFLY polynomials for torus knots, in agreement with the Rosso and Jones formula. We show that our A-model construction is mirror symmetric to the B-model analysis of Brini, Eynard and Mari\~no. Comparing to the recent proposal by Aganagic and Vafa for knots on $S^3$, we demonstrate that the disk amplitude of the A-brane associated to any knot is sufficient to reconstruct the entire B-model spectral curve. Finally, the construction of toric Lagrangian A-branes is generalized to other local toric Calabi-Yau geometries, which paves the road to study knots in other three-manifolds such as lens spaces.

\smallskip
\Date{December 2012}


\newsec{Introduction}

Since its discovery, the remarkable relation \WittenJones~between the observables of Chern-Simons gauge theory on the one hand and knot invariants on the other hand has opened a new window to both physics and mathematics. The realization of Chern-Simons theory as an open A-model topological string theory \WittCSA, and its relation to a large $N$ dual closed A-model topological strings \GV\ have further enriched the relationship between the realm of topological strings and knot theory. According to \OV, insertion of Lagrangian probe branes promotes the large $N$ duality picture to incorporate knots and links in the geometry as well. More precisely, one first associates to each knot $K$ in $S^{3}$ a Lagrangian brane $L_{\CK}$ in the target space $T^{*}S^{3}$. After the large $N$ transition, one ends up with a priori new Lagrangian brane ${\tilde{L}}_{\CK}$ in the resolved conifold geometry. The conjecture of \OV\ predicts that the open-string partition function of the resolved conifold in the presence of $\tilde{L}_{\CK}$ encodes the invariants of the knot $\CK$ in $S^{3}$. Incorporating M-theory into the picture \GVM, one of the striking consequences of \OV\ is the explanation of the integrality of the Jones polynomial.

The equivalence between knot invariants and the topological string amplitudes has thoroughly been verified for the framed unknot and the Hopf link (for a compact review see \MRev~and references therein). Recently, a lot of effort has been devoted to verify the conjecture of \OV~beyond the case of unknot. The complication to complete this task is twofold. First, for each knot $\CK$, one needs to construct the Lagrangian submanifolds $L_{\CK}$ and $\tilde{L}_{\CK}$ precisely. Initial progresses to address this issue for the case of complex algebraic knots have appeared in \refs{\LMV,\Taub,\Kosh}. However the delicate construction of the Lagrangian cycles $L_{\CK}$ and $\tilde{L}_{\CK}$, corresponding to algebraic knots, has recently been performed in \DSV. The second complication in verifying the conjecture of \OV\ is actually to compute the open-string amplitudes corresponding to the Lagrangian cycle $\tilde{L}_{\CK}$, with the current techniques of topological string theory. In certain situations where one has extra symmetries in the geometrical setup, one might be able to perform the computation of the topological amplitudes. It has been shown in \DSV\ that for the case of torus knots, the existence of a torus action allows one to employ the localization techniques of \refs{\KL,\LS} to compute the open-string amplitudes of the constructed Lagrangian cycle to reproduce the HOMFLY polynomial of torus knots, in agreement with the Rosso and Jones formula \RJ.

In virtue of mirror symmetry, in order to avoid the complications of dealing with Lagrangian cycles, torus knots have also been studied from the B-model point of view \refs{\BEM,\AVtwo}. In the approach by Brini, Eynard and Mari\~no \BEM, the key role is played by the $SL(2,\IZ)$ transformation acting on a choice of local B-model coordinates, so as to arrive for each torus knot $\CK_{r,s}$ at a spectral curve. The relevant $SL(2,\IZ)$ action maps to the $SL(2,\IZ)$ transformation in the Chern-Simons theory that generates the torus knot $\CK_{r,s}$ from an unknot \BEM. Since all these spectral curves are effectively of genus zero, applying the topological recursion techniques \refs{\EO,\BKMP} are feasible, which enables the authors of \BEM\ to rederive the colored HOMFLY polynomials of torus knots $\CK_{r,s}$ in a closed form. However, this approach is tailored for torus knot $\CK_{r,s}$ and does not seem to generalize to more general knots on $S^3$. Aganagic and Vafa have recently put forward a proposal \AVtwo, which --- at least in principal --- provides for a framework to arrive at the B-model spectral curves for any knot on $S^3$. These spectral curves are suitable to capture directly the moduli space of the Lagrangian A-brane associated to any knot ${\cal K}$ on $S^3$. The authors derive these spectral curves from the set of HOMFLY polynomials colored with symmetric representations, and it is observed in \refs{\AVtwo,\NgJX}~that these curves correspond to the augmentation polynomials of the knot $\CK$ on $S^3$ \Ng.\foot{The correspondence between spectral curves and augmentation polynomials has also recently been explored for torus knots $\CK_{2,2p+1}$ in \FujiNX.}

In this note, we aim for implementing the symplectic transformation considered in \BEM~ directly in the A-model picture. To achieve this, we first construct a Lagrangian cycle in the deformed conifold, which has the topology $T^{2}\times\IR$. A torus knot $\CK_{r,s}$ on $S^{3}$ can now be obtained by performing a symplectic transformation on the $(1,0)$ cycle of torus of the constructed Lagrangian. After the large $N$ transition, for generic values of the open moduli, we find a toric Lagrangian in the resolved conifold with the same topology. We then degenerate the dual cycle of the knot and this results in placing the Lagrangian on one leg of the toric skeleton. In order to compute the open topological amplitudes associated to the new Lagrangian cycle, we apply the topological vertex \AKMV, and eventually reproduce the colored HOMFLY polynomial of the corresponding torus knot. We argue that the topological vertex approach to torus knots $\CK_{r,s}$ on $S^3$ also generalizes to study knots on more general three-manifolds such as lens spaces.

Using local mirror symmetry, we recognize that the constructed Lagrangian A-branes are mirror-symmetric to the spectral curve of \BEM, which describes torus knots on $S^3$. We compare these spectral curves of \BEM\ to the spectral curves appearing in \AVtwo. The former spectral curves are feasible to apply topological recursion relations \refs{\EO,\BKMP}, since they are effectively of genus zero. However, as we explain --- compared to the latter spectral curves in \AVtwo\ --- the spectral curves of \BEM\ are of rather of auxiliary type, which makes it difficult to generalize them to arbitrary knots $\CK$ on $S^3$. We argue that in order to construct the spectral curves proposed in \AVtwo, which in principal directly describe the quantum moduli space of the Lagrangian A-branes $L_\CK$ associated to any knot $\CK$, it is sufficient to know the disk amplitude of the A-brane $L_\CK$. We demonstrate this procedure by explicitly constructing spectral curves of torus knots from their disk amplitudes. As proposed in \refs{\AVtwo,\NgJX} the calculated spectral curves are in agreement with the
augmentation polynomials of the described torus knots on $S^3$ \Ng.

We proceed as follows. In section~2, we construct the Lagrangian cycles describing torus knots in the deformed and resolved conifold geometries. In section~3, we apply the topological vertex to compute the open-string amplitudes of our proposed toric  Lagrangian in the resolved conifold, and rederive the Rosso and Jones formula --- that is to say the colored HOMFLY polynomial of a torus knot $\CK_{r,s}$ as a decomposition in terms of the usual quantum dimensions. In section~4, we first briefly review the basic aspects of the B-model approach delivered in \BEM. Guided by mirror symmetrywe realize that the new toric Lagrangian cycle corresponding to a torus knot $\CK_{r,s}$ in the resolved conifold is mirror-symmetric to the spectral curve derived in \BEM. We show how the mirror curves of torus knots obtained in \BEM\ and \AVtwo\ are related, and we demonstrate that the spectral curve of \AVtwo\ can be constructed merely by the knowledge of the genus zero disk instantons. In section~5, we show how our construction generalizes to other toric geometries, which are in particular helpful for studying torus knots in lens spaces. We conclude in section 6. Appendix A concerns a technical detail about Adams operation, and we explicitly represent the spectral curve of the torus knot $\CK_{3,5}$ in appendix~B.

\newsec{A-model} \seclab\secAmodel
In this section, we describe in the torus knots on $S^3$ in terms of non-compact Lagrangian probe branes on the deformed conifold $T^*S^3$. In particular we are interested in the effect of $SL(2,\IZ)$ transformations acting upon these probe branes as the deformed conifold undergoes a large $N$ transition to the resolved conifold $\CO(-1)\oplus\CO(-1)\rightarrow\IP^1$. From this process, we conclude that any non-compact probe Lagrangian on $T^*S^3$ --- which is associated to a torus knot on $S^3$ --- enjoys a description in terms of a multi-sheeted cover of the toric non-compact Lagrangian on $\CO(-1)\oplus\CO(-1)\rightarrow\IP^1$, which is associated to the unknot. Due to the multi-sheeted-covering property, we argue that any torus knot may effectively be described in terms of the toric non-compact Lagrangian of the unknot with an appropariatly chosen and apparently fractional framing.

\subsec{Lagrangian Branes from Torus Knots} \subseclab\secSLags
In order to describe torus knots on $S^3$ in the resolved conifold geometry $\CO(-1)\oplus\CO(-1)\rightarrow\IP^1$, we briefly recall their origin in the deformed conifold geometry $T^*S^3$ via the large $N$ transition \refs{\OV,\LMV}. Given a knot $\CK$ on $S^3$ we construct a three-dimensional (non-compact) special Lagrangian $\CK\times\IR^2$ by extending the knot on $\CK$ into the cotangent bundle of $S^3$ such that it becomes a special Lagrangian of topology $\CK\times\IR^2$ \refs{\OV,\LMV,\Taub}. The relevant Wilson loop observables arise at the intersection $\CK$ of a stack of $N$ branes on $S^3$ and the probe brane on $\CK\times\IR^2$.

To describe a large $N$ transition, we move the probe brane along the cotangent direction of $S^3$, such that this displaced probe brane does not intersect the zero section of $T^*S^3$ anymore \refs{\LMV,\DSV}. As a consequence, in the large $N$ transition to the resolved geometry --- which arises from an extremal transition that is topologically described by a surgery operation in the vicinity of the zero section of $T^*S^3$ --- the displaced probe brane remains unaffected, and therefore we arrive at a Lagrangian brane of topology $\CK\times \IR^2$ in $\CO(-1)\oplus\CO(-1)\rightarrow\IP^1$. This probe brane is also displaced from the zero section of the resolved conifold fibration.

To describe the torus knot $\CK_{r,s}$ with the topological vertex formalism, we employ a similar strategy. We view the base $S^3$ in $T^*S^3$ as a fibration of a two-torus over an interval, where the two one-cycles, which correspond to the $(1,0)$- and $(0,1)$-cycles, degenerate at the respective boundaries of the interval as depicted in~\lfig\LargeN. The central fiber of such a fibration is extended in $T^*S^3$ to a (non-compact) Lagrangian $\CL$ of topology $T^2 \times\IR$ as depicted in \lfig\LargeN. The torus $T^2$ of $\CL$ comes with a symplectic basis $(\ac,\bc)$ of one cycles generating $H_1(T^2,\IZ)$. The one-cycle represented by the homology class $\tilde\ac=r\,\ac + s\,\bc$ describes a torus knot $\CK_{r,s}$ in $S^3$, which we use to construct a probe brane $\CK_{r,s}\times\IR^2$. As discussed in ref.~\AKV, we can alternatively arrive at such a probe brane $\CK_{r,s}\times \IR^2$ by picking a component that arises from degenerating the Lagrangian $\CL$ in such a way that a one-cycle $\tilde\bc$ dual to $r\,\ac + s\,\bc$ becomes trivial.

Note that for the local Calabi-Yau threefold $T^*S^3$, we focus here on a situation where a pair of symplectic dual one-cycles of $T^2$ --- e.g., a $(1,0)$- and $(0,1)$-cycle --- degenerates at the boundary of the base interval, respectively. This construction generalizes to other sets of generating pairs of $H_1(T^2,\IZ)$, i.e., two generators that do not necessarily furnish a symplectic basis with respect to the intersection pairing. We return to this aspect in detail in Section~5\yyy{}, where we explain that such a generalization is suitable to describe torus knots on Lens spaces.

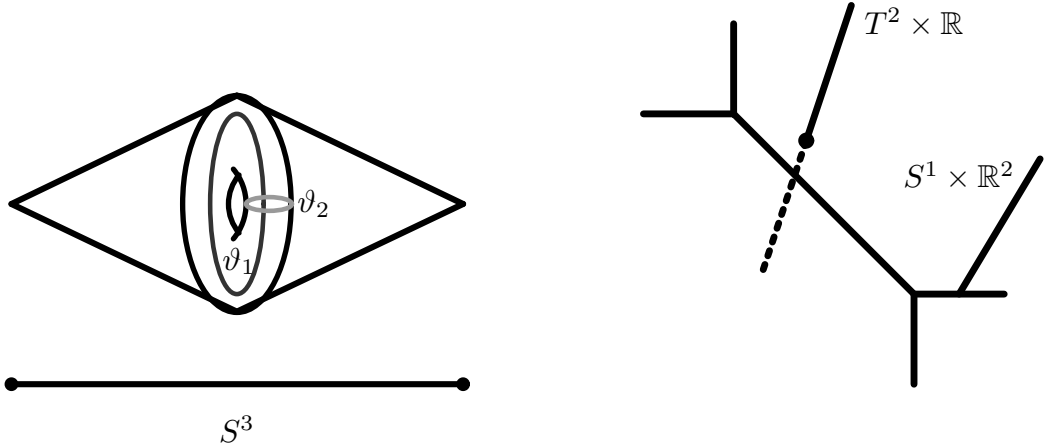
\figboxinsert\LargeN{The left drawing depicts the three sphere as a $T^2$-fibration over the interval, which is the compact part of the deformed conifold $T^*S^3$. In this fibration the symplectic dual one cycles $\ac$ and $\bc$ degenerate at the left and right boundaries, respectively. The central $T^2$-fiber gives rise to a central fiber Lagrangian brane of topology $T^2\times\IR$. The right drawing shows the toric skeleton of the resolved conifold $\CO(-1)\oplus\CO(-1)\rightarrow \IP^1$ together with the Lagrangian brane of topology $T^2\times\IR$, which upon a large $N$ transition is associated to the central fiber brane in $T^*S^3$. Its open-string moduli can be tuned such that it degenerates into two components of topology $S^1\times\IR^2$ with one component also shown in the right diagram.}{
\DrawDiag{
\drawdim cm \linewd 0.06
\move(0 6)
\move(4 3) \lellip rx:0.6 ry:1.2
\move(3.5 3) \larc r:0.6 sd:320 ed:40
\move(4.4 3) \larc r:0.5 sd:140 ed:220
\linewd 0.07
\move(1.5 3) \lvec(4 4.2)
\move(1.5 3) \lvec(4 1.8)
\move(6.5 3) \lvec(4 4.2)
\move(6.5 3) \lvec(4 1.8)
\move(1.5 1) \lvec(6.5 1)
\linewd 0.05
\move (1.5 1) \lcir r:0.05
\move (6.5 1) \lcir r:0.05
\setgray 0.2
\move(4 3) \lellip rx:0.3 ry:1
\setgray 0.6
\move(4.35 3) \lellip rx:0.25 ry:0.08
\setgray 0 \linewd 0.08
\move(11.5 2) \lvec(12.5 2) \move(11.5 2)
\lvec(11.5 1) \move(11.5 2)
\rlvec(-2 2) \rlvec(-1 0)
\move(9.5 4) \rlvec(0 1)
\move(10.3 3.7) \rlvec(0.5 1.5)
\move(10.3 3.7) \lcir r:0.05
\move(12 2) \rlvec(0.9 1.5)
\textref h:C v:C \htext(11.5 5){$T^{2}\times\IR$}
\textref h:C v:C \htext(12 3.3){$S^{1}\times\IR^{2}$}
\textref h:C v:C \htext(4.025 2.35){$\ac$}
\textref h:C v:C \htext(4.85 3){$\bc$}
\textref h:C v:C \htext(4 0.5){$S^{3}$}
\lpatt(0.06 0.1)
\move(10.3 3.7) \rlvec(-0.5 -1.5)
\move(14.25 0.2)
}}

Returning to the local Calabi-Yau threefold $T^*S^3$, in order to analyze torus knots in the resolved conifold geometry, we perform again a large $N$ transition. Now we displace the Lagrangian $\CL$ into the cotangent direction of $T^*S^3$ such that it again remains topologically unaffected by the transition, and we obtain a special Lagrangian cycle $T^2\times \IR$ in the resolved conifold geometry as depicted on the right hand side of \lfig\LargeN. In the resolved conifold the resulting special Lagrangian cycle $T^2\times \IR$ enjoys a toric description \AKV. Explicitly, the bulk geometry may be described by the usual symplectic quotient construction via the charge vector $P=(1,1,-1,-1)$ and the associated moment map equation
\eqn\Pcharge{|X_{1}|^{2}+|X_{2}|^{2}-|X_{3}|^{2}-|X_{4}|^{2}={\rm Re}\,t\ ,}
where $t$ is the complexified K\"ahler modulus of the resolved conifold geometry. Then the special Lagrangian cycle $T^2\times\IR$ arises from the two charges $p^{1}=(0,1,-1,0)$ and $p^{2}=(0,0,-1,1)$ inducing the momentum map equations
\eqn\slagcharge{\eqalign{|X_{2}|^{2}-|X_{3}|^{2}&={\rm Re}\,c_{1}\ ,\cr -|X_{3}|^{2}+|X_{4}|^{2}&={\rm Re}\,c_{2}\ .}}
Here $c_1$ and $c_2$ are the complexified cycle moduli of $T^2\times\IR$ with the identifications $c_1 \sim c_1 + 2\pi i$ and $c_2 \sim c_2 + 2\pi i$. The periodic imaginary parts of the moduli $c_1$ and $c_2$ are identified with the phases of the coordinates $X_2$ and $X_4$, respectively, which in turn parametrize the two one-cycles $\ac$ and $\bc$ of the torus of the described special Lagrangian cycle. For ${\rm Re}\,c_{1}>0$ and ${\rm Re}\,c_{2}>0$ the coordinates $X_2$ and $X_4$ never vanish on the Lagrangian brane. However, in the limit $c_1\rightarrow 0$ the Lagrangian $T^2\times\IR$ degenerates into two components of topology $S^1\times\IR^2$ \AKV, which are the special Lagrangian branes $S^1\times\IR^2$ associated to the unknot $S^1\simeq\CK_{1,0}$. One of the Lagrangian components $S^1\times\IR^2$ embedded into the resolved conifold $\CO(-1)\oplus\CO(-1) \rightarrow \IP^1$ becomes \AV
\eqn\UnknotSLag{\varphi_{\CK_{1,0}}(\theta;R,\chi) = \left( \sqrt{{\rm Re}\,t+{\rm Re}\,c_2+R^2}\,e^{-i\theta}, R\,e^{i\chi},R\,e^{-i\chi},\sqrt{{\rm Re}\,c_2+R^2}\,e^{i\theta} \right) \ , }
where $\theta$ is the angular coordinate of $S^1\simeq\CK_{1,0}$ and $(R,\chi)$ are polar coordinates of $\IR^2$.

The constructed Lagrangian \UnknotSLag\ preserves the two $S^1$-symmetries associated to both the degenerating and the non-degenerating one-cycles. These two real symmetries rotate the phases of the complex coordinates $X_2$ and $X_4$ independently, they arise from compact parts of two toric $\IC^*$-symmetries of the toric variety $\CO(-1)\oplus\CO(-1) \rightarrow \IP^1$, and they are essential in applying the topological vertex to the discussed geometry.  For general torus knots $\CK_{r,s}$ with $r>0,s>0$ and $r,s$ co-prime, we need to degenerate a one-cycle $\tilde\bc=p\,\ac+q\,\bc$ (with $rq-ps=1$) of the Lagrangian cycle $T^2\times\IR$ dual to $\tilde\ac=r\,\ac+s\,\bc$, so as to obtain a special Lagrangian brane component $\CK_{r,s}\times\IR^2$. Unfortunately, a degeneration of the cycle $\tilde\bc$ while keeping the cycle $\tilde\ac$ non-degenerate is not possible in a $\IC^*$-symmetry compatible manner, as the cycles $\tilde\ac$ and $\tilde\bc$ are associated to the respective phases of the real coordinates $\sqrt{r|X_2|^2+s|X_4|^2}$ and $\sqrt{p|X_2|^2+q|X_4|^2}$, which in a toric degeneration vanish simultaneously.

Therefore, we employ an alternative construction. First, we act with an $SL(2,\IZ)$ transformation
\eqn\SLtwoZ{ M= \pmatrix{r & s \cr p & q}\in SL(2,\IZ)\qquad,\qquad rq-ps=1\ .}
on the brane charges $\pmatrix{p_1 \cr p_2}\mapsto \pmatrix{\tilde p_1 \cr \tilde p_2}=M \pmatrix{p_1 \cr p_2}$ and arrive at the charges $\tilde p_1=(0,r,-r-s,s)$ and $\tilde p_2=(0,p,-p-q,q)$ for the momentum map equations
\eqn\newlagchar{\eqalign{r |X_{2}|^{2}-(r+s)|X_{3}|^{2}+s|X_4|^2&={\rm Re}\,\tilde c_{1}\ ,\cr p|X_2|^2-(p+q)|X_{3}|^{2}+q|X_{4}|^{2}&={\rm Re}\,\tilde c_{2}\ ,}}
with the moduli $\tilde c_1 \sim \tilde c_1 + 2\pi i$ and $\tilde c_2 \sim \tilde c_2 + 2\pi i$. For generic values ${\rm Re}\,\tilde c_{1}$ and ${\rm Re}\,\tilde c_{2}$ the resulting Lagrangian has topology $T^2\times \IR$ and retracts to the compact torus $T^2$. For $X_3=0$ on the Lagrangian~\newlagchar\ we readily identify the holonomies of $\tilde c_1$ and $\tilde c_2$ with the phases of the real coordinates $\sqrt{r|X_2|^2+s|X_4|^2}$ and $\sqrt{p|X_2|^2+q|X_4|^2}$. Thus, due to the $SL(2,\IZ)$-transformation the two relevant one-cycles become manifest, because the imaginary part of $\tilde c_1$ captures the holonomy around the knot $\CK_{r,s}$ associated to the one-cycle $\tilde\ac=r\,\ac + s\,\bc$, whereas $\tilde c_2$ describes the holonomy around the dual one-cycle $\tilde\bc=p\,\ac +q\,\bc$.

For the description of a probe brane $\CK_{r,s}\times\IR^2$ of the torus knot we could try to describe a non-toric deformation of $T^2\times\IR$ such that the dual one-cycle $\tilde\bc$ degenerates. However, in order to arrive at a formulation that is more suitable for the topological vertex, we view $T^2\times\IR$ as an auxiliary Lagrangian cycle, which allows us to construct a probe brane of topology $\tilde\CL=\{p\}\times\CK_{r,s}\times\IR^2$, where $p$ is a point on the one-cycle $\tilde\bc$ while the knot $\CK_{r,s}$ is identified with the one-cycle~$\tilde\ac$. Alternatively, we can think of this probe brane as arising from two T-dualities along the cycle $\tilde\bc$ and a suitable second non-compact direction. For the obtained probe brane $\tilde\CL$ the imaginary part of $c_1$ describes the holonomy of $\tilde\CL$ along the knot $\CK_{r,s}$, while the imaginary part of $c_2$ parameterizes the position of the point $p$ on the dual one-cycle $\tilde\bc$.

To analyze the probe brane $\tilde\CL$ of interest, it is convenient to undo the $SL(2,\IZ)$ action on the level of the charges $\tilde p_1$ and $\tilde p_2$. As result we are required to act with the same inverse $SL(2,\IZ)$ transformation on the moduli $\tilde c_1$ and $\tilde c_2$. Then we arrive at
\eqn\newcharge{\eqalign{|X_{2}|^{2}-|X_{3}|^{2}&={\rm Re}\,\hat c_{1}\ ,\cr -|X_{3}|^{2}+|X_{4}|^{2}&={\rm Re}\,\hat c_{2}\ ,}}
with
\eqn\newc{ \pmatrix{ \hat c_1 \cr \hat c_2 } = M^{-1} \pmatrix{ c_1 \cr c_2 }=\pmatrix{ q\,\tilde c_1 - s\,\tilde c_2 \cr  -p\,\tilde c_1 + r\,\tilde c_2 }\ . }
To obtain in terms of the new open parameters $\hat c_1$ and $\hat c_2$ a probe brane of the correct topology $\CK_{r,s}\times\IR^2$ from \newcharge, we take the limit $\hat c_1=0$, which yields
\eqn\crels{ \hat c_2 = {1\over s} \tilde c_1 \ , \qquad \tilde c_2= {q\over s} \tilde c_1 \ . }
Let us now pause to interpret these relations. Since the holonomy of $\tilde c_1$ associated to the knot $\CK_{r,s}$ has periodicity $2\pi$, the non-zero open-string parameter $\hat c_2$ has periodicity $\hat c_2 \sim \hat c_2 + 2\pi i\,s$. As a result --- at least on the level of topology --- we can interpret the probe brane $\tilde\CL$ as a $s$-sheeted cover of the unknot $\CK_{1,0}$. This observation reflects the derivation of the Rosso-Jones formula in the Chern-Simons derivation in ref.~\RJ, in which one first considers an $s$ cable of the unknot ({\it i.e.} one winds $s$ times around the cycle $\ac$ of the boundary torus) and acquires $s$ powers of the holonomy operator. In the next subsection, we will see that the modified periodicity of the modulus also affects its apparent framing of the Lagrangian \newcharge. The second relation \crels\ simply states that the position of the point $p$ parameterized by $\tilde c_2$ is correlated with the open-string modulus $\tilde c_1$ accordingly.

\subsec{Framing} \subseclab\secFraming
As we realized in the previous subsection, in order to describe the Lagrangian probe brane associated to the torus knot $\CK_{r,s}$, we constructed an $s$-sheeted cover of the toric Lagrangian associated to the unknot. Since the imaginary part of the open moduli $\tilde c_1$ and $\hat c_2$ parametrize the Wilson loops around the torus knot $\CK_{r,s}$ and the unknot $\CK_{1,0}$, respectively, we identify --- as reflected in the identification \crels\ --- the Wilson loop around $\CK_{r,s}$ with an $s$-fold Wilson loop around the unknot $\CK_{1,0}$. Thus, we arrive at the idenfication
\eqn\Ascale{ \oint_{\CK_{r,s}} \!\!\! \tilde A\,=\, s\,\oint_{\CK_{1,0}}\!\!\!\hat A \ ,}
where $\hat A$ and $\tilde A$ denote gauge connections along the respective Wilson loops.

In order to see the effect of the identification~\Ascale, let us review how the framing factor arises \GMM. In the computation of the vacuum expectation value of the holonomy operator around a knot $\CK$, one encounters the expectation value of the composite operator~$(\oint_{{\cal K}}A)^{2}$
\eqn\cotor{{\rm Tr}(t^{a}t^{b})\oint_{{\cal K}}dx^{\mu}\int^{x}dy^{\nu}\langle A^{b}_{\nu}(y)A^{a}_{\mu}(x)\rangle={2\pi i\over k}{\rm dim}G\,c_{2}(G)\,\varphi({\cal K})\ ,}
where $G$ is the Chern-Simons gauge group with generators $t^{a}$ and $c_2(G)$ is its second Casimir operator. In \cotor, $\varphi({\cal K})$ is by definition the cotorsion of ${\cal K}$ and it is given by
\eqn\vphi{\varphi({\cal K})={1\over 4\pi}\oint_{{\cal K}}dx^{\mu}\oint_{{\cal K}}dy^{\nu}\,\epsilon_{\mu\nu\rho}{(x-y)^{\rho}\over |x-y|^{3}}\ .}
Although the cotorsion is well-defined and finite, it is, however, a metric dependent quantity which is not invariant under deformations of the knot ${\cal K}$. Therefore, the cotorsion cannot be a topological invariant of ${\cal K}$. This problem can be cured due to the existence of a degree of freedom \WittenJones~in defining the composite operator $(\oint_{{\cal K}}A)^{2}$. The solution involves a companion knot ${\cal K}_{f}$ which is constructed via a unit normal vector to the actual knot ${\cal K}$
\eqn\normal{x^{\mu}(t)\mapsto y^{\mu}(t)=x^{\mu}(t)+\varepsilon\, n^{\mu}(t)\ ,}
where $n^{\mu}(t)$ is a normal vector field ($|n(t)|=1$) and $0<\varepsilon\ll 1$. Now, instead of dealing with $\varphi({\cal K})$, we can work with the following quantity which is constructed via the companion knot ${\cal K}_{f}$
\eqn\fphi{\eqalign{\varphi_{f}({\cal K})&=\lim_{\varepsilon\rightarrow0}{1\over 4\pi}\int_{0}^{1}ds\int_{0}^{1}dt\,\epsilon_{\mu\nu\rho}\,\dot{x}^{\mu}(s)\dot{y}^{\nu}(t)
{(x(s)-y(t))^{\rho}\over |x(s)-y(t)|^{3}}\cr
&=\lim_{\varepsilon\rightarrow0}{1\over 4\pi}\oint_{{\cal K}}dx^{\mu}\oint_{{\cal K}_{f}}dy^{\nu}\,\epsilon_{\mu\nu\rho}{(x-y)^{\rho}\over |x-y|^{3}}=\lim_{\varepsilon\rightarrow0}\,{\rm lnk}({\cal K},{\cal K}_{f})\ .}}

In the second line of \fphi, we recognize that the Gauss integral is nothing except the linking number between the actual knot ${\cal K}$ and its companion ${\cal K}_{f}$. However, the linking number
\eqn\link{{\rm lnk}({\cal K},{\cal K}_{f})={1\over 4\pi}\oint_{{\cal K}}dx^{\mu}
\oint_{{\cal K}_{f}}dy^{\nu}\,\epsilon_{\mu\nu\rho}{(x-y)^{\rho}\over |x-y|^{3}}\ ,}
is an integer and is independent of $\varepsilon$. Therefore, we can remove $\lim_{\varepsilon\rightarrow0}$ from \fphi, and the cotorsion $\varphi_{f}({\cal K})$ simply becomes
\eqn\linking{\varphi_{f}({\cal K})={\rm lnk}({\cal K},{\cal K}_{f})\ ,}
which is now a topological invariant. This integer which is defined via the choice of a companion knot is known as the framing of the knot ${\cal K}$.

Now, let us return to our discussion. Relating the holonomy of the torus knot $\CK_{r,s}$ to the holonomy of the unknot $\CK_{1,0}$ according to \Ascale, we immediately conclude that the framings are also related accordingly, namely
\eqn\frscale{\varphi_{f}(\CK_{r,s})\,=\, s^{2}\, \varphi_{f}(\CK_{1,0}) \ . }
%

%
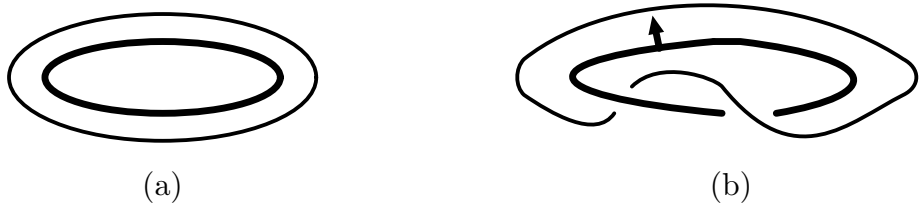
\figboxinsert\Framing{Two examples of framing. The actual knot ${\cal K}$ is displayed by the thick line and its companion ${\cal K}_{f}$ by the thin line.}{
\DrawDiag{
\drawdim cm \linewd 0.08
\move (4 4)
\move(2 2)
\lellip rx:1.3 ry:0.4 \linewd 0.04 \lellip rx:1.7 ry:0.7
\linewd 0.08 \move(8.2 1.6)
\clvec (6 1.8)(6 2.2)(8.1 2.4)
\linewd 0.04
\arrowheadsize l:0.2 w:0.18 \arrowheadtype t:F
\linewd 0.08
\move(7.5 2.32)
\ravec(-0.08 0.36)
\move(8.1 2.4)
\clvec (8.2 2.4)(8.3 2.4)(8.4 2.4)
\clvec (10 2.2)(10 1.8)(8.8 1.6)
\linewd 0.04 \move(7 1.6)
\clvec (6.8 1.3)(6.3 1.6)(6 1.8)
\clvec (5.9 1.9)(5.9 2.1)(6 2.2)
\clvec (7 3)(9 3)(10.2 2.2)
\clvec (10.4 2.1)(10.4 1.9)(10.2 1.8)
\clvec (9.6 1.5)(9 0.9)(8.2 1.9)
\clvec (8.1 2)(7.5 2.2)(7.2 1.9)
\textref h:C v:C \htext(2 0.8){(a)}
\textref h:C v:C \htext(8.3 0.8){(b)}
\move(0 0.4)
}}

For the torus knot $\CK_{r,s}$ in $S^{3}$, there exists a natural framing that is given by the linking number $\varphi^{\rm nat}_f(\CK_{r,s})=rs$. This is also the framing to be used here, since the Rosso-Jones formula~\RJ\ for torus knots applies to this natural framing. As a consequence, using the natural framing for the torus knot, in terms of the framing of the unknot, we arrive at the following identification
\eqn\fracframe{\varphi^{\rm nat}_f(\CK_{1,0}) \,=\, {1\over s^2}\, \varphi^{\rm nat}_f(\CK_{r,s}) \,=\, {r\over s} \ . }

Let us pause to interpret this result. Our analysis yields that the Lagrangian probe brane $\CK_{r,s}\times \IR^2$ on $T^*S^3$ associated to the torus knot $\CK_{r,s}$ on $S^3$ enjoys an alternative description in terms of the $s$-fold cover of the toric Lagrangian for the unknot. In order to arrive at the most general possible integral framing for this $s$-fold covering space, we allow ${1\over s}$-fractional framings for the toric Lagrangian of the unknot. In particular, the natural framing of the torus knot $\CK_{r,s}$ maps to the fractional framing \fracframe\ of the unknot $\CK_{1,0}$. This fractional framing appears as a fractional twist of the torus knot operator in the Chern-Simons theory \LLR, and fractional framings have also already appeared in the context of open-string A-model localization calculations \refs{\DFone,\DFtwo}. In our context, the fractional framing of the unknot becomes integral again, once we consider Wilson loops of $\CK_{r,s}$ that are described in the physically relevant $s$-fold covering space of the toric Lagrangian for the unknot $\CK_{1,0}$.

\newsec{Topological Vertex and the Colored HOMFLY Polynomial} \seclab\secVertex
In this section, we proceed with the computation of open-string amplitudes in the resolved conifold associated to the probe brane Lagrangian for torus knot $\CK_{r,s}$. These amplitudes calculate the HOMFLY polynomial of $\CK_{r,s}$. In the previous section we have given an equivalent description for the probe brane of $\CK_{r,s}$ in terms of toric Lagrangians on the resolved conifold. In this section, we use this result to put forward an extended application for the topological vertex that is suitable to calculate such open-string amplitudes.

In relating the probe brane Lagrangian for the torus knot $\CK_{r,s}$ to the unknot, we related in \crels\ the open-string moduli of the probe brane Lagrangian to the open-string modulus of the toric Lagrangian \newcharge\ with the explained adjusted periodicity of the imaginary part. In the topological vertex, the open-string modulus has been absorbed in the holonomy matrix, and taking the adjusted periodicity \crels\ into account amounts to the replacement
\eqn\holsc{\CV_{\CK_{1,0}} \,=\, {\tilde\CV}^{1\over s}_{\CK_{r,s}} \ , }
where ${\tilde\CV}_{\CK_{r,s}}$ is the holonomy matrix for the disk ending on the Lagrangian cycle describing the torus knot $\CK_{r,s}$. To ease notation, we also denote this holonomy matrix simply by $\CV$. Since the holonomy matrix now appears in the open-string amplitudes with fractional powers, this implies that the full open-string amplitude contains fractional winding. However, only amplitudes with integral windings are physical and we need to extract those integral winding amplitudes.

In order to extract the integral winding contributions, we recall that the trace of a holonomy matrix ${\cal U}$ in the winding basis is given by \AKMV
\eqn\trhol{{\rm Tr}_{\vec{k}}{\cal U}=\prod_{j=1}^{\infty}\Big({\rm tr}\,{\cal U}^{j}\Big)^{k_{j}}\ .}
The open-string partition function corresponding to the insertion of the special Lagrangian cycle \newcharge~on an outer leg of the toric skeleton of the resolved conifold (as depicted in \lfig\Znu) has the following structure
\eqn\openamp{Z=\sum_{\vec{k}}{1\over z_{\vec{k}}}\,Z_{\vec{k}}\,{\rm Tr}_{\vec{k}}{\cal V}_{\CK_{1,0}}\ ,}
where $z_{\vec{k}}=\prod_{j}k_{j}!\,j^{k_{j}}$. Here, we have written the sum in the winding basis over all possible windings $\vec{k}=(k_{1},k_{2},k_{3},\ldots)$. Thus in order to achieve integral windings in terms of the holonomy matrices  $\CV=\tilde{\CV}_{\CK_{r,s}}$, we consider only winding states whose winding vector has the following form
\eqn\kvec{\matrix{ & j=1&  j=2 &  & j=s & j=s+1&  & j=2s &  & \cr
\vec{k}^{(s)}=( & 0\, , & 0\, , & \ldots & k_{1}, &0\, ,& \ldots & k_{2}, & \ldots)\ .}}
Note that the entries at positions $j=\ell s$ for $\ell=1,2,3,\ldots$ in the above winding vector~$\vec{k}^{(s)}$ represent the physical states, as they give rise to an integral holonomy matrix. The remaining entries of the winding vector have to be set to zero in order to avoid non-integral holonomy matrices for unphysical states. This reflects the fact that the Lagrangian for the torus knot $\CK_{r,s}$ is described by the $s$-sheeted cover of the toric Lagrangian for the unknot. Therefore, a holonomy one-cycle on the unknot Lagrangian lifts only to closed loop on the relevant $s$-sheeted covering Lagrangian, if it winds a multiple of $s$ times. This explains geometrically the selection rule for physical states appearing in winding vector \kvec.

The interpretation of the B-model $SL(2,\IZ)$ action in the A-model leads to open string amplitudes with fractional windings and framings. Open string amplitudes have been calculated in the A-model using equivariant localization with an insertion
of descendent operators which mimicks the insertion of loop operators in the B-model~\Br. This technique also applies to (partially) resolved orbifold geometries~\BC~and can be used to calculate and interpret the fractional framing
and winding contributions encountered above within an orbifold geometry~\BCprog.

Alternatively, instead of taking the sum over all windings in \openamp, we consider the sum to run only over the states that have the form \kvec
\eqn\Zint{Z_{\rm int}=\sum_{\vec{k}^{(s)}}{1\over z_{\vec{k}^{(s)}}}\,Z_{\vec{k}^{(s)}}\,{\rm Tr}_{\vec{k}}{\cal V}\ ,}
with ${\rm Tr}_{\vec{k}^{(s)}}{\cal V}_{\CK_{1,0}}={\rm Tr}_{\vec{k}}{\cal V}$ and where the winding vector $\vec{k}$ is obtained by deleting all zeros in the positions where $j$ is not a multiple of $s$, i.e., $\vec{k}=(k_{1},k_{2},k_{3},\ldots)$. Thus in \Zint\ we run the sum over the winding vector $\vec{k}$ instead of $\vec{k}^{(s)}$ (by deleting zeros). However, this introduces a combinatorics factor $s^{|\vec{k}|-1}$ ($|\vec{k}|=\sum_{j}k_{j}$ is the total number of holes), which was first considered in \LM. Therefore, the integer winding part of the open-string partition function is rewritten to
\eqn\Zinttwo{Z_{\rm int}={1\over s}\sum_{\vec{k}}{s^{|\vec{k}|}\over z_{\vec{k}^{(s)}}}\,Z_{\vec{k}^{(s)}}\,{\rm Tr}_{\vec{k}}{\cal V}={1\over s}\sum_{\vec{k}}{1\over z_{\vec{k}}}\,Z_{\vec{k}^{(s)}}\,{\rm Tr}_{\vec{k}}{\cal V}\ ,}
where we used $z_{\vec{k}^{(s)}}=s^{|\vec{k}|}z_{\vec{k}}$.

We can now use the transformation rules of the topological vertex to express the ingredients of the above expression in the representation basis. To do this, we first recall that \AKMV
\eqn\windTr{{\rm Tr}_{\vec{k}}{\cal V}=\sum_{|\mu|=\ell(\vec{k})}\chi_{\mu}(C_{\vec{k}})\,{\rm Tr}_{\mu}{\cal V}\ ,}
where the sum runs over all partitions whose length is equal to the total winding number $\ell(\vec{k})=\sum_{j}jk_{j}$ corresponding to winding vector $\vec{k}$. In \windTr, $\chi_{\mu}(C_{\vec{k}})$ is the character of the symmetric group $S_{\ell(\vec{k})}$ for the conjugacy class $C_{\vec{k}}$, in the representation specified by the Young tableau $\mu$. We can also express the amplitude $Z_{\vec{k}^{(s)}}$ in the representation basis. This amplitude corresponds to the insertion of the Lagrangian cycle on one external leg, and is colored with a nontrivial representation. Under this change of basis, we obtain
\eqn\Zrep{Z_{\vec{k}^{(s)}}=\sum_{|\nu|=s\cdot\ell(\vec{k})}\chi_{\nu}
(C_{\vec{k}^{(s)}})\,Z_{\nu}\ .}
We have now expressed all ingredients of $Z_{\rm int}$ in the representation basis. Substituting \windTr~and \Zrep~into \Zinttwo, we find
\eqn\Zintthr{\eqalign{Z_{\rm int}&={1\over s}\sum_{\mu}\sum_{|\nu|=s|\mu|}\Big(\sum_{\vec{k}}{1\over z_{\vec{k}}}\chi_{\mu}(C_{\vec{k}})\chi_{\nu}(C_{\vec{k}^{(s)}})\Big)Z_{\nu}\,{\rm Tr}_{\mu}{\cal V}\cr
&={1\over s}\sum_{\mu}\sum_{|\nu|=s|\mu|}c_{\mu,s}^{\nu}\,Z_{\nu}\,{\rm Tr}_{\mu}{\cal V}\ ,}}
where the sum over $\mu$ runs over all partitions. The integer coefficients $c_{\mu,s}^{\nu}$ are the coefficients of Adams operation. For more details about Adams operation, the corresponding coefficients and their properties see appendix A.

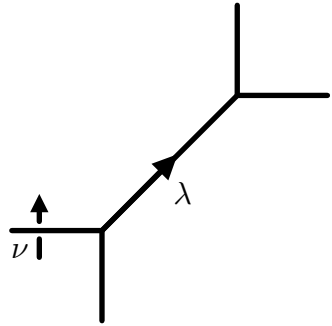
\figboxinsert\Znu{The open-string amplitude $Z_{\nu}$ in ${\cal O}(-1)\oplus {\cal O}(-1)\rightarrow\IP^{1}$.}{
\DrawDiag{
\drawdim cm \linewd 0.06
\move(2 6)
\move(2 2) \lvec(2 3) \move(2 3) \lvec(1 3) \move(2 3) \rlvec(1.5 1.5) \rlvec(1 0) \move(3.5 4.5) \rlvec(0 1) \move(1.3 2.7) \rlvec(0 0.2) \move(1.3 3.1)
\arrowheadsize l:0.2 w:0.2 \arrowheadtype t:F \avec(1.3 3.4) \move(2 3)
\arrowheadsize l:0.28 w:0.26 \avec(2.84 3.84) \textref h:C v:C \htext(1.1 2.75){$\nu$} \textref h:C v:C \htext(2.9 3.4){$\lambda$}
}}

Now, let us compute the open-string amplitude $Z_{\nu}$. This is an amplitude corresponding to insertion of a Lagrangian on one of the outer legs of the skeleton of the resolved conifold. As argued in \fracframe, the corresponding framing factor for this Lagrangian must be the fractional number ${r\over s}$.\foot{Note that, since we are only considering multiples of $s$-winding states as physical, the framing factor actually becomes integral on the level of physical states.} Under a change of framing by $f$ units, the corresponding open-string amplitude is multiplied by the factor $e^{2\pi if h_{\mu}}=Q^{{f\over 2}|\mu|}q^{{f\over 2}\kappa_{\mu}}$ where $h_{\mu}$ is conformal weight of the primary field associated with $\mu$. Since the first factor $Q^{{f\over 2}|\mu|}$ is an overall multiplicative factor, it is often absorbed in the definition of the holonomy matrix \MV. However, we would like to have this factor explicitly represented, and we do not absorb it in the definition of ${\cal V}$.
Using the rules of the topological vertex \AKMV, the open-string amplitude $Z_{\nu}$ is given by
\eqn\Znu{\eqalign{Z_{\nu}&=\sum_{\lambda}(-Q)^{|\lambda|}Q^{{r\over 2s}|\nu|}q^{{r\over 2s}\kappa_{\nu}}C_{\bullet\lambda\nu}
C_{\bullet\bullet\bar{\lambda}}\cr
&=Q^{{r\over 2s}|\nu|}q^{{r\over 2s}\kappa_{\nu}}\sum_{\lambda}(-Q)^{|\lambda|}q^{\kappa_{\lambda}\over 2}s_{\bar{\nu}}(q^{-\rho})s_{\lambda}(q^{-\rho-\bar{\nu}})s_{\lambda}(q^{-\rho})\cr
&=Q^{{r\over 2s}|\nu|}q^{{r\over 2s}\kappa_{\nu}}s_{\bar{\nu}}(q^{-\rho})\sum_{\lambda}(-Q)^{|\lambda|}s_{\bar{\lambda}}
(q^{-\rho})s_{\lambda}(q^{-\rho-\bar{\nu}})\cr
&=Q^{{r\over 2s}|\nu|}q^{{r\over 2s}\kappa_{\nu}}s_{\bar{\nu}}(q^{-\rho})\prod_{i,j=1}^{\infty}\big(1-Q\,
q^{i+j-\bar{\nu}_{j}-1}\big)\ ,}}
where $\bar{\nu}$ and $\bar{\lambda}$ are the conjugate representations of $\nu$ and $\lambda$ respectively. In \Znu, we have used the Schur function representation of the topological vertex  \ORV\ and a similar computation has been already carried out in \GIKV~for the case of the Hopf link. In the last line of \Znu, we have used the Schur function identity \MacD
\eqn\idone{\sum_{\lambda}s_{\lambda}(x)s_{\bar{\lambda}}(y)=\prod_{i,j}(1+x_{i}y_{j})\ .}

The open-string partition function $Z_{\nu}$ is the unnormalized amplitude. To make contact with knot invariants, we should consider the normalized amplitudes. This means that in order to get the normalized open-string amplitude, we must divide $Z_{\nu}$ by the closed-string partition function $Z_{\varnothing}$
\eqn\Zpart{Z_{\varnothing}=\prod_{i,j=1}^{\infty}\big(1-Q\,q^{i+j-1}\big)\ .}
Let $Z_{\nu}^{\rm norm}$ denote the corresponding normalized open-string amplitude. Using \Znu~and \Zpart, we find
\eqn\Znorm{\eqalign{Z_{\nu}^{\rm norm}\equiv{Z_{\nu}\over Z_{\varnothing}}&=Q^{{r\over 2s}|\nu|}q^{{r\over 2s}\kappa_{\nu}}s_{\bar{\nu}}(q^{-\rho})\prod_{(i,j)\in\nu}\big(1-Q\,q^{j-i}\big)\cr
&=Q^{{r\over 2s}|\nu|}q^{{r\over 2s}\kappa_{\nu}}\,{\rm dim}_{q}\nu\ ,}}
where ${\rm dim}_{q}\nu$ is the quantum dimension of the representation given by the partition $\nu$. In a similar manner, we can normalize the integer winding part of the full open-string partition function
\eqn\Zintnorm{Z_{\rm int}^{\rm norm}\equiv {Z_{\rm int}\over Z_{\varnothing}}={1\over s}\sum_{\mu}\sum_{|\nu|=s|\mu|}c_{\mu,s}^{\nu} Q^{{r\over 2s}|\nu|}q^{{r\over 2s}\kappa_{\nu}}\,{\rm dim}_{q}\nu\, {\rm Tr}_{\mu}{\cal V}\ .}
Apart from an irrelevant numerical factor of ${1\over s}$,\foot{As reviewed in the next section, a similar numerical prefactor appears in the mirror B-model derivation \BEM.} the coefficient of ${\rm Tr}_{\mu}{\cal V}$ corresponds to the HOMFLY polynomial of an $(r,s)$ torus knot colored with representation $\mu$
\eqn\HOMFLY{W_{\mu}({\cal K}_{r,s})=\sum_{|\nu|=s|\mu|}c_{\mu,s}^{\nu}\,e^{2\pi i{r\over s}h_{\nu}}\,{\rm dim}_{q}\nu\ ,}
where $h_{\nu}$ is the corresponding conformal weight associated with representation $\nu$ in the WZW model.
\par
Thus, in summary we have shown how to apply the topological vertex to accommodate for the $s$-fold cover of the toric Lagrangian for the unknot, which yields a suitable description for the torus knot $\CK_{r,s}$. It correctly reproduces the Rosso-Jones identity, which expresses the (colored) HOMFLY polynomial of the torus knot $\CK_{r,s}$ as a superposition of the usual quantum dimensions.


\newsec{Comparison to the B-model via Local Mirror Symmetry}

The aim of this section is to study the mirror symmetric B-model geometry for knots $\CK$ on $S^3$. To this end we first review the approach taken by Brini, Enyard and Mari\~no \BEM\ towards torus knots $\CK_{r,s}$ on $S^3$. We show that their construction is mirror-symmetric to our A-model derivation of toric A-branes presented in section~2 and section~3. These leads us towards a comparison with the recent proposal by Aganagic and Vafa \AVtwo, where spectral curves for knots on $S^3$ are derived. We show that these spectral curves can straight-forwardly be obtained from the disk amplitude of the associated A-brane $L_\CK$. We demonstrate this feature explicitly and derive new B-model spectral curves for various torus knots and links.


\subsec{Torus Knots and $SL(2,\IZ)$ Transformations in the B-model} \subseclab\secSLtwoCurves
In this section, we first review the B-model treatment of torus knots as discussed in \BEM. The key point of \BEM\ is to identify the $SL(2,\IZ)$ transformation that generates a torus knot $\CK_{r,s}$ from the unknot $\CK_{1,0}$ on $S^3$ with the B-model $SL(2,\IZ)$ action which rotates the open-string moduli. From Chern-Simons theory point of view, the $SL(2,\IZ)$ transformation acting on the unknot $\CK_{1,0}$ has a natural lift on the Hilbert space of states of the exact quantized Chern-Simons theory. Using the explicit form of torus knot operators \LLR, one conveniently derives the Wilson loop expectation values of Chern-Simons theory along torus knots. In this way, Chern-Simons theory realizes the Rosso and Jones formula \refs{\Stev,\BEM}, which expresses the HOMFLY polynomial of a torus knot $\CK_{r,s}$ in terms of the usual quantum dimensions.

The B-model setup of \BEM\ furnishes the local mirror geometry of the resolved conifold $\CO(-1)\oplus\CO(-1)\rightarrow\IP^1$, which reads \refs{\AV,\AKV}
\eqn\Mirrorcurve{u\,v\,=\, H(\UU,\VV;Q) \ , \quad H(\UU,\VV;Q)\,=\,1+\UU+\VV+Q\,\UU\VV \ ,}
in terms of the local coordinates $(u,v)\in \IC^2$ and $(\UU,\VV)\in \IC^{*}\times\IC^{*}$ and the complex structure modulus $Q$. A point $(\UU,\VV)\in\IC^*\times\IC^*$ on the mirror curve given by $H(\UU,\VV;Q)=0$ describes a non-compact B-brane at $u=0$, which is mirror to a Lagrangian A-brane for the unknot $\CK_{1,0}$. One can therefore understand the curve $H(\UU,\VV;Q)=0$ as the open string moduli space of this B-brane, which can propagate on this curve.

For non-compact B-branes dual to Lagrangian A-branes, the holomorphic Chern Simons action is localized and yields the superpotential
\eqn\Superpotential{W(\UU,Q)\,=\,\int_{\UU_*}^\UU \lambda \ , }
with a reference point $\UU_*$ on the curve $H(\UU,\VV;Q)=0$ and with the differential
\eqn\SuperDiff{\lambda\,=\,\log(\VV) { \dd\UU \over \UU} , }
which is interpreted by mirror symmetry as the generating function for the disk instantons ending on the Lagrangian A-brane.

Introducing a B-brane in the geometry corresponds to considering a point on this mirror curve. We can assume that the point is specified by the choice of $\UU$ and then \Mirrorcurve\ specifies $\VV$ as
\eqn\Mirrorsol{\VV\,=\,-{1+\UU\over 1+Q\,\UU} \ .}
The function $\VV(\UU)$ is the generating function for the genus zero contribution of the Wilson loop expectation values of the unknot $\CK_{1,0}$ in the standard framing
\eqn\expan{-\log \VV(\UU)\,=\,\sum_{n\geq0}\langle{\rm Tr}\,\CV_{\CK_{(1,0)}}^{n}\rangle_{g=0}\,\UU^{n}\ ,}
where $\CV_{\CK_{(1,0)}}$ captures the holonomy matrix around the unknot $\CK_{1,0}$.

Based on the results of \AKV\ in order to study the unknot in a non-standard framing, it is necessary to extract a different generating function $\YY(\XX)$ from the mirror curve $H(\UU,\VV)=0$. This is conveniently achieved by introducing new local coordinates $(\XX,\YY)\in\IC^*\times\IC^*$ arising from a $T$-transformation that acts upon the local coordinates $(\UU,\VV)$ according to $\UU\to \XX=\UU\VV^{f}$ and $\VV\to\YY=\VV$. Then mirror curve is given in parametric form as a function of $\UU$ by
\eqn\framing{\XX\,=\,\UU\left({-1-\UU\over 1+Q\,\UU}\right)^{f} \ , \qquad \YY\,=\,\VV\,=\,-{1+\UU\over 1+Q\,\UU}\ .}
Analogously to \expan\ eliminating $\UU$ in favor of $\XX$ yields the generating function $\YY(\XX)$ that generates the genus zero contribution of the Wilson loop expectation values of the unknot with $f$ units of framing.

In order to study an $(r,s)$ torus knot, the authors of \BEM\ generalize such a local coordinate transformation to a general $SL(2,\IZ)$ transformation acting upon the local coordinates $(\UU,\VV)$ of the mirror curve \BEM
\eqn\SLtran{\pmatrix{\UU\cr\VV}\to\pmatrix{\XX\cr\YY}\,=\,\pmatrix{\UU^{s}\,\VV^{r}\cr \UU^{q} \,\VV^{p}}\ , \quad \pmatrix{s & r \cr q & p}\in SL(2,\IZ) \ . }
One can eliminate $\UU$ and $p$ from equation $\YY=\UU^s\VV^p$ using $\XX=\UU^s\VV^r$ and the unimodularity of elements of $SL(2,\IZ)$ to arrive at
\eqn\Supo{\log\YY\,=\, {q\over s}\log\XX - {1\over s} \log\VV(\XX) \ ,}
where the relation $\VV(\XX)$ is implicitly defined  by
\eqn\XvU{\XX\,=\,\UU\left(-{1+\UU \over 1+Q\UU}\right)^{r\over s}\ , \qquad \VV\,=\,-\left(1+\UU\over 1+Q\UU\right)\  ,}
or equivalently by
\eqn\XvV{\widetilde H_{r,s}(\XX,\VV;Q)\,=\,(1+Q\VV)^s\XX-(-1-\VV)^s\VV^r\ . }
The explicit function
\eqn\logV{\log\VV(\XX)\,=\,-\sum_{n=1}^\infty W_n (Q)\,\XX^{n\over s}}
is most easily obtained by inverting \XvU\ in favor of $\UU(\XX^{1\over s})$ and inserting this into \XvU. The coefficients can be written in a closed form as \BEM
\eqn\Wn{W_n(Q)={1\over n!}\sum_{l=0}^n (-1)^{n+l}  \left(n\atop l\right) Q^{(1-l-n{r \over  s})} \prod_{j=-l+1}^{m s -1- l} ( m r -j)\ .}
The authors of \BEM\ argue that the non-analytic first term in \Supo\ is an irrelevant classical term, which can be dropped, and they interpret \XvV\ with the idenfitication  $\VV=\YY^{-s}$  as the mirror curve. Given the fractional powers of the open modulus $\XX$ in \logV, it is clear that $\log\YY$ in \Supo\ cannot be the derivative of the disk instanton generating appearing in the differential \SuperDiff. As a consequence --- unlike the original curve \Mirrorcurve\ associated to the unknot $\CK_{1,0}$ or the ones describing the change of framing according to \framing\ --- the curve $\widetilde H_{s,r}(\XX,\VV;Q)=0$ cannot directly be interpreted as the moduli space of the brane mirror dual to the special Lagrangian A-brane for the torus knot $\CK_{r,s}$. For this reason we consider \XvV\ as an auxiliary curve. Note in particular that this auxiliary curve is not symmetric in $(r,s)$. If $r$ and $s$ are exchanged, different powers  in  $\log\VV(\XX)$ have to be discarded to make contact with the disk superpotential \Superpotential\ associated to the torus knot $\CK_{r,s}$.

Comparing \XvU\ with \framing, we conclude that $f$ is identified with the fractional value $f={r\over s}$. This is consistent with the Chern-Simons
derivation of torus knot invariants, in which ${r\over s}$ appears as a fractional twist of torus knot operators. Therefore, one concludes \BEM\ that the general $SL(2,\IZ)$ degree of freedom in the choice of local coordinates of the B-model is identified with the general $SL(2,\IZ)$ action that generates a torus knot from an unknot.

Nevertheless, the relation~\Supo\ does encode the information of disks or equivalently the genus zero expectation values $\left\langle {\rm tr} {\cal V}^n_{ {\cal K}_{r,s}}
\right\rangle_{g=0}$ of the torus knots ${\cal K}_{r,s}$, if one restricts the sum in \logV\ to integer powers of $\XX$.

The remodelling principle of \BKMP\ interprets  the mirror curve \Mirrorcurve\ as the spectral curve of a matrix model, where the differential $\lambda$ becomes the filling fractions of this matrix model. From the spectral curve and the one form $\lambda$ all amplitudes of the matrix model can be reconstructed by the topological recursion \EO .

Moreover, it has been shown that the closed string amplitudes are invariants under the symplectic transformation \SLtran\ in this formalism. This fits well with the interpretation of the $SL(2,\IZ)$ transformed curves as mirror curves of the framed torus knots or links invariants, as in the $A$-model geometry the closed string geometry is invariantly given by $T^* S^3$.

As mentioned in the case of integer framing one has $\YY=\VV$ and the curve \XvV\ can be immediately be identified with the spectral curve associated to the framed unknot. In the general case the spectral curve is \XvV\ supplemented by relation $\VV=\YY^{-s}$. In both cases we get a highly degenerate higher genus curve with nodal singularities so that its geometric genus is zero. This implies that the non-vanishing period integrals are just the ones of the orginal \Mirrorcurve\ of the unknot, in accordance with the expectation that the closed string sector does not change.

Interpreting this spectral curve given by the meromorphic functions $\XX(\UU)$ and $\YY(\UU)$ parametrized by the affine coordinate $\UU$ on $\IP^1$, one can define recursively the meromorphic differentials $\omega_{g,h} \dd\UU_1\ldots \dd\UU_h$ on the $h$-fold cartesian product of the spectral curve \EO, which calculate all open-string amplitudes in various parametrizations of the open-string modulus, in particular also for the local coordinate $\XX$. More precisely, the differentials $\omega_{g,h} \dd\UU_1\ldots \dd\UU_h$ are defined by the topological recursions relations arising from the Bergmann kernel \EO, which reads for spectral curves of geometric genus zero
\eqn\Bergmann{ B(\UU_1,\UU_2)\dd\UU_1\dd\UU_2\,=\,{\dd\UU_1\dd\UU_2 \over (\UU_1-\UU_2)^2}\ , }
and the corresponding recursion kernel
\eqn\Recursionkernel{K(\UU_1,\UU_2)\,=\,-{\XX(\UU_2)\int^{\UU_2}_{\bar\UU_2} B(\UU_1,\UU) \dd\UU \over  2\XX'(\UU_2)(\log\YY(\UU_2)- \log\YY(\bar\UU_2)) } \  ,}
where $\bar{\UU}_{2}$ is the conjugate (under the projection of the mirror curve to $\IP^{1}$) of $\UU_{2}$ near a branch point of the mirror curve. That is to say, from the spectral curves alone all differentials $\omega_{g,h} \dd\UU_1\ldots \dd\UU_h$, and hence all information about colored HOMFLY polynomials for the torus knots $\CK_{r,s}$ can be calculated in this way \BEM.

\subsec{Local Mirror Symmetry for Torus Knots}

In this section, we show that we can appropriately translate the $SL(2,\IZ)$ transformation of the B-model that we described in the previous section to the A-model via the dictionary of mirror symmetry. As a result, we obtain that the proposed Lagrangian for torus knots that we introduced in section 2.1 is in agreement with Lagrangian cycle we find by the virtue of mirror symmetry.

In the A-model picture, the bulk geometry, after the large $N$ transition, is the resolved conifold which is described by \Pcharge~in the toric language. The mirror curve \Mirrorcurve,  associated with the unknot, is associated to the Lagrangian brane \slagcharge~in the A-model. This Lagrangian lands on one of the outer legs of the toric skeleton, and has the topology $S^{1}\times\IR^{2}$ (this equivalently implies that for this brane $c_{1}=0$). According to the philosophy of \BEM, we are now supposed to perform an $SL(2,\IZ)$ transformation on the $\IC^{*}$-coordinates of the mirror curve \Mirrorcurve. On the other hand, the dictionary of mirror symmetry tells us how the A- and B-model geometries are related, and hence enables us to trace the effect of the B-model symplectic transformation on the A-model open/closed geometry. The open/closed mirror A- and B-model geometries are related via mirror symmetry \refs{\HV,\AV}~in the following way
\eqn\Qcons{\prod_{i}y_{i}^{P_{i}}=Q\ ,\quad \prod_{i}y_{i}^{\,p_{i}^{\alpha}}=e^{c_{\alpha}}\ ,\qquad \alpha\in\{1,2\}\ ,}
where $P$ and $p^{\alpha}$ are the toric charges of the bulk and brane geometries respectively. In \Qcons, $y_{i}$ are the local coordinates of the B-model, and $Q=e^{-t}$ is the closed-string modulus.

Using \Qcons, we would now like to translate the B-model symplectic transformation \SLtran~to the A-model setup. Recall that introducing a torus knot $\CK_{r,s}$ amounts to performing a symplectic transformation on the local coordinates of the B-model \SLtran. Before we implement this transformation, let us first choose a local patch of coordinates. We choose the patch where $y_{2}\neq0$ and introduce the affine (inhomogeneous) coordinates as
\eqn\Ycoor{\alpha={y_{3}\over y_{2}}\quad,\quad \beta={y_{4}\over y_{2}}\quad,\quad \gamma={y_{1}\over y_{2}}\ .}
Without loss of generality, we assume that the mirror curve is described by coordinates $(\alpha,\beta)$. According to \SLtran, in order to introduce an $(r,s)$ torus knot, we are supposed to perform a symplectic transformation on these local coordinates
\eqn\sly{\pmatrix{\alpha \cr \beta}\mapsto\pmatrix{\hat{\alpha}\cr \hat{\beta}}=\pmatrix{\alpha^{r}\,\beta^{s}\cr \alpha^{p}\,\beta^{q}}\ .}
Using the inverse transformation, we can express the old coordinates in terms of the new ones in a one-to-one fashion as $\alpha=\hat{\alpha}^{q}\hat{\beta}^{-s}$ and $\beta=\hat{\alpha}^{-p}\hat{\beta}^{r}$.
\par
As pointed out in \BEM, an $SL(2,\IZ)$ transformation of the local coordinates of the mirror curve is a symmetry of the bulk geometry of the B-model. In order to accomplish this, we transform the $\gamma$ coordinate in a way that it preserves the bulk geometry. The most general transformation on $\gamma$ takes the following form
\eqn\yone{\gamma=\hat{\gamma}^{a}\hat{\alpha}^{b}\hat{\beta}^{c}\ ,}
where $a$, $b$, and $c$ are some integers. Now, the first equation of \Qcons, written in the patch $y_{2}\neq0$,
\eqn\QconY{\gamma=Q\, \alpha \,\beta\ ,}
can be expressed in terms of the new variables $\hat{\alpha}$, $\hat{\beta}$, $\hat{\gamma}$. Rewriting this in terms of the homogenous coordinates $\hat{y}_{i}$, we find
\eqn\conitran{\hat{y}_{1}^{a}\,\hat{y}_{2}^{-(a+b+c+p-q+s-r)}\,\hat{y}_{3}^{b+p-q}\,
\hat{y}_{4}^{c+s-r}=Q\ .}
From \conitran, we can now read off the new toric charge vector of the bulk geometry $\hat{P}=(a,-a-b-c-p-s+q+r,b+p-q,c+s-r)$. Requiring that the new bulk geometry must be again the resolved conifold, we have $\hat{P}=P$. This determines the values of $a$, $b$, and $c$
\eqn\abc{a=1\qquad,\qquad b=q-p-1\qquad,\qquad c=r-s-1\ .}
Plugging these values in \yone\ and using \sly, we find that the transformation of the $Y_{1}$ coordinate
\eqn\yoneplug{\hat{\gamma}=\gamma\,\alpha^{r+p-1}\,\beta^{s+q-1}\ .}
\par
We can now proceed for the A-brane  and perform the same game with Lagrangian charges $p^{1}$ and $p^{2}$. However, we notice that we do not have any degree of freedom left in transforming the coordinates. Therefore, when we perform the transformation on Lagrangian charges, we should find a new set of charges describing a new Lagrangian. As we will see, this Lagrangian is the one which produces the invariants of the torus knot $\CK_{r,s}$ and coincides with the Lagrangian cycle proposed in section 2.1. Let us start with the first charge $p^{1}=(0,1,-1,0)$. Working in the same patch, we obtain from the second equation of \Qcons
\eqn\pone{\alpha^{-1}=e^{c_{1}}\ .}
To obtain the Lagrangian cycle associated with the transformed mirror curve, we express the old variables in terms of the new ones, and rewrite the result in terms of the homogeneous coordinates. We then have
\eqn\ponenew{\hat{y}_{2}^{q-s}\,\hat{y}_{3}^{-q}\,\hat{y}_{4}^{s}=e^{c_{1}}\ .}
Therefore, the first Lagrangian charge transforms to $\hat{p}^{1}=(0,q-s,-q,s)$. We can now follow the same procedure for the second Lagrangian charge $p^{2}=(0,0,-1,1)$. Again from the second equation of \Qcons\ we find
\eqn\ptwo{\alpha^{-1}\beta=e^{c_{2}}\ .}
Similar to the previous Lagrangian charge, if we rewrite \ptwo\ in terms of the new variables and express the result in terms of the homogeneous variables, we gain
\eqn\ptwonew{\hat{y}_{2}^{p+q-r-s}\,\hat{y}_{3}^{-p-q}\,\hat{y}_{4}^{r+s}=e^{c_{2}}\ .}
Hence the second new Lagrangian charge is found to be $\bar{p}^{2}=(0,p+q-r-s,-p-q,r+s)$. Instead of working with $\bar{p}^{2}$, we can equivalently work with the charge $\hat{p}^{2}=\hat{p}^{1}-\bar{p}^{2}$. Therefore, the two new Lagrangian charges are found to be $\hat{p}^{1}=(0,q-s,-q,s)$ and $\hat{p}^{2}=(0,r-p,p,-r)$. In terms of the A-model coordinates, this implies
\eqn\newlagcharII{\eqalign{&(q-s)|\hat{X}_{2}|^{2}-q|\hat{X}_{3}|^{2}+
s|\hat{X}_{4}|^{2}={\rm Re}\,\tilde{c}_{1}\ ,\cr
& (r-p)|\hat{X}_{2}|^{2}+p|\hat{X}_{3}|^{2}-r|\hat{X}_{4}|^{2}
={\rm Re}\,\tilde{c}_{1}-{\rm Re}\,\tilde{c}_{2}\ .}}
We notice that the new Lagrangian which is described by $\hat{p}^{1}$ and $\hat{p}^{2}$ is a special Lagrangian cycle. This Lagrangian submanifold is the mirror of the transformed mirror curve of the B-model, which is supposed to carry the information about an $(r,s)$ torus knot. In order to analyze the topology of the Lagrangian cycle \newlagcharII, we perform the same trick as in section 2.1 by considering another linear combinations of the Lagrangian charges, and instead, pushing the transformation into the open moduli space. To see this explicitly, we consider the following Lagrangian charges
\eqn\ptilde{\eqalign{&\tilde{p}^{1}=r\,\hat{p}^{1}+s\,\hat{p}^{2}\ ,\cr
&\tilde{p}^{2}=p\,\hat{p}^{1}+q\,\hat{p}^{2}\ .}}
Finally, we arrive at the following D-term equations describing the Lagrangian cycle
\eqn\newlagtwo{\eqalign{&|\hat{X}_{2}|^{2}-|\hat{X}_{3}|^{2}
=r{\rm Re}\,\tilde{c}_{1}+s({\rm Re}\,\tilde{c}_{1}-{\rm Re}\,\tilde{c}_{2})\ ,\cr
&|\hat{X}_{2}|^{2}-|\hat{X}_{4}|^{2}=p{\rm Re}\,\tilde{c}_{1}+q({\rm Re}\,\tilde{c}_{1}-{\rm Re}\,\tilde{c}_{2})\ .}}
Instead of the second equation of \newlagtwo, we can equivalently use the difference between the two equations of \newlagtwo
\eqn\newlagtwo{\eqalign{|\hat{X}_{2}|^{2}-|\hat{X}_{3}|^{2}&=r{\rm Re}\,\tilde{c}_{1}
+s({\rm Re}\,\tilde{c}_{1}-{\rm Re}\,\tilde{c}_{2})\ ,\cr
-|\hat{X}_{3}|^{2}+|\hat{X}_{4}|^{2}&=(r-p){\rm Re}\,\tilde{c}_{1}+(s-q)({\rm Re}\,\tilde{c}_{1}-{\rm Re}\,\tilde{c}_{2})\ .}}
This shows that instead of working with the original charges $\hat{p}^{1}$ and $\hat{p}^{2}$, we can work with the untransformed charge vectors $p^{1}$ and $p^{2}$, but we need to perform the following $SL(2,\IZ)$ transformation in the open moduli space
\eqn\openmod{\pmatrix{\tilde{c}_{1}\cr \tilde{c}_{2}-\tilde{c}_{1}}\mapsto \pmatrix{\hat{c}_{1}\cr \hat{c}_{2}}=\pmatrix{r & -s\cr r-p & q-s}\pmatrix{\tilde{c}_{1}\cr \tilde{c}_{2}-\tilde{c}_{1}}\ .}
This, in a sense, is the passive way of the transformation. Instead of performing the symplectic transformation directly on the Lagrangian submanifold, the $SL(2,\IZ)$ tranformation allows us to keep the Lagrangian unaffected, but we perform the transformation in the moduli space of the Lagrangian.
\par
Now in this passive form of the transformation, it is easier to analyze the topology of the Lagrangian cycle. For generic values of the open moduli, the Lagrangian , as pointed out in section 2.1, will have the topology $T^{2}\times\IR$. However, for special values of the open moduli, the Lagrangian cycle will have the topology of $S^{1}\times\IR^{2}$. The Lagrangian \newlagtwo~(or equivalently \newlagcharII) is the A-model mirror of the transformed mirror curve which describes the torus knot. Therefore, it is required that the Lagrangian \newlagcharII~lands on one of the outer legs of the toric skeleton. This is easily achieved by setting $\hat{c}_{1}=0$. This eliminates $\tilde{c}_{2}$ in favor of $\tilde{c}_{1}$. Then, the open modulus of the Lagrangian  $\hat{c}_{2}$ is then found as
\eqn\lagmodulus{\hat{c}_{2}={1\over s}\tilde{c}_{1}\ .}
The above equation is the analog of \crels\ that we found by directly in the A-model. Again, we may interpret the resulting Lagrangian as an $s$-sheeted cover of the Lagrangian associated with the unknot. As discussed in section~\secFraming, the rescaling of the open modulus \lagmodulus\ has also an effect on the framing factor by making it  fractional with respect to the Lagrangian associated to the unknot. Altogether we realize that the Lagrangian cycles for torus knots $\CK_{r,s}$ obtained by local mirror symmetry agrees with the proposed Lagrangian A-branes of section~\secSLags.

\subsec{Disk Amplitudes and Spectral Curves}

From the discussion in section \secSLtwoCurves, it becomes clear that all information of torus knots can be very efficiently extracted by the method in \BEM\  using the $SL(2,\IZ)$ symmetry \SLtran, which is the mirror symmetric construction of the A-model approach described in section~\secAmodel\ and section~\secVertex. However, this $SL(2,\IZ)$ construction results in a curve $\widetilde H_{r,s}(\XX,\VV;Q)$ in eq.~\XvV\ that is of rather auxiliary nature. For instance, as discussed the relevant expansion for disk amplitudes \logV, contains apart from the disk instantons additional redundant non-analytic information. This also manifests itself that for a given torus knot $\CK_{s,r}$, we can actually assign (at least) two distinct auxiliary curves $\widetilde H_{r,s}(\XX,\VV;Q)$ and $\widetilde H_{s,r}(\XX,\VV;Q)$, which encode the same disk amplitudes in the described fashion.

Due to the prominent role of $SL(2,\IZ)$ transformations in the construction of \BEM\ --- which are motivated by $SL(2,\IZ)$ actions transforming the unknot $\CK_{1,0}$ into the torus knot $\CK_{r,s}$ as reflected in the the Rosso-Jones identity \RJ\ --- this approach restricts to torus knots. In general one would hope that the true B-model curve $H_\CK(\alpha,\beta)=0$ can be geometrically constructed by mirror symmetry as the (quantum) moduli space of the special Lagrangian brane $L_\CK$, which in turn is determined for any knot $\CK$ on $S^3$.

Then a suitable phase of the special Lagrangian brane $L_\CK$ determines on the spectral curve $H_\CK(\alpha,\beta)=0$ an expansion point, which we choose in the following to reside at $\alpha=0$, such that the differential
\eqn\Disks{ \lambda \,=\, \log \VV(\UU) {d\UU \over \UU} \ , }
yields directly the quantum superpotential \Superpotential\ without the need to eliminate any ambiguous non-analytic terms in the instanton expansion.

Note that this disk amplitude is associated to a matrix model, for which the large $N$ limit of the loop equation of the matrix model yields the classical spectral curve $H_\CK(\alpha,\beta)=0$ \refs{\DV,\MM,\BKMP}. The Eynard-Oratin recursion relation \EO\ allows us to retrieve the full large $N$ expansion from the classical spectral curve, which in turn contains the full information about all colored HOMFLY polynomials. We will argue below that the knowledge of the disk amplitude associated to the brane $L_\CK$ is equivalent to the information of the classical spectral curve $H_\CK(\alpha,\beta)=0$, which in turn allows us to extract (at least in principal) any colored HOMFLY polynomial of $\CK$.

A more conceptual way to phrase the relation between the classical spectral curve and the finite $N$ or quantum geometry was proposed in \ADKMV, which is then applied to spectral curves $H_{r,s}(\alpha,\beta)=0$ of torus knots $\CK_{r,s}$ in \AV. The relation between the classical geometry and the quantum geometry containing the coupling $g_s\sim 1/N$ is given by promoting the spectral curve $H(\UU,\VV)=0$ of the mirror A-brane $L$ with $\UU=e^x$ and $\VV=e^p$ to a quantum operator
\eqn\Quantumcurve{\widehat H( e^{\hat x}, e^{\hat p}) \left| \Psi_L \right\rangle = 0 \ , }
which annihilates the quantum state $\left| \Psi_L \right\rangle$. Here $\hat x$ and $ \hat p$ are canonically conjugated operators
\eqn\Commutator{[\hat p, \hat x]=g_s\ .}
which in position space are identified with $\hat x=x$ and $\hat p=g_s \partial_x$. The topological open-string partition function $Z_L(g_s,x)$ is then viewed as the wave function in position space
\eqn\OpenPart{ Z_L(g_s,x) \,=\, \left\langle\,\hat x\, | \, \Psi_L \, \right\rangle \ , }
and the classical geometry is recovered in the WKB approximation as
\eqn\semiclass{Z_L(g_s,x)\sim \exp \left({1\over g_s}\int p(x) {\rm d} x\right)\ .}

For A-branes $L_\CK$ describing knots on the deformed conifold $T^*S^3$, the open-string partition function $Z_{L_\CK}(g_s,x,Q)$ encodes the HOMFLY polynomials colored with symmetric representations \refs{\AVtwo}
\eqn\OpenPar{ Z_{L_\CK}(g_s,x,Q) \,=\, \sum_{n=0}^{+\infty} \left\langle {\rm tr}_{S_n} \CU \right\rangle_{\CK} e^{n x } \ , }
where $S_n$ corresponds to representation of the totally symmetric Young tableau with $n$ boxes. In particular the disk contribution ${1\over g_s}$ to the partition function $Z_{L_\CK}(g_s,x,Q)$ --- and hence the contributions at order ${1\over g_s}$ of the HOMFLY polynomials colored with symmetric representations --- contains the entire disk amplitude, which in turn gives rise in the planar large $N$ limit to the quantum superpotential \Superpotential\ according to
\eqn\wkb{\log\VV(\UU) = \alpha\,{d\over d\alpha}\,\lim_{g_s\to 0} \  g_s \log \left( \sum_{n=0}^\infty \left\langle {\rm tr}_{S_n} \CU \right\rangle_{\CK} \UU^n \right) \ . }
Given a normal ordering prescription one can use \Quantumcurve\ to determine $Z(g_s,x,Q)$ as was demonstrated for genus zero curves in \ADKMV. However, it was also realized there that this is much more involved for higher genus curves, which one obtains for a generic knot. The formalism of Eynard and Orantin seems more developed even in this case \EO.

Our philosophy towards the spectral curve $H_\CK(\UU,\VV)=0$ for a given knot $\CK$ is the following: Given the disk amplitude associated to a knot $\CK$ --- that is to say the quantum corrected superpotential \Superpotential\ --- we can unambiguously construct the (irreducible) spectral curve $H_\CK(\UU,\VV)=0$, which then encodes the entire higher genus information, which is --- at least in principal --- accessible by the formalism of Eynard and Orantin \EO. As the set of all HOMFLY polynomials colored with symmetric representations reduce in the planar large $N$ limit to the entire disk amplitude \refs{\OV,\LMV}, this information is more than sufficient to construct the spectral curves $H_\CK(\UU,\VV,Q)=0$. This has been demonstrated in \AVtwo, where the authors also construct the spectral curve associated to the figure eight knot (and which does not fall in the class of torus knots).

Let us start with the first step of finding the classical curve for the torus knot $\CK_{r,s}$. Here the information about the disks is contained in the mod $s$ coefficients of \Wn . Using the actual genus zero disk amplitude we will reconstruct an algebraic curve $H(\UU,\VV;Q)=0$, which one can view as the actual moduli space of the brane $L_{\CK_{r,s}}$. Let us define therefore
\eqn\bva{\VV(\UU)=\exp \left(-\sum_{n=1}^\infty W_{sn}(Q) \UU^n \right)\ .}
The method for finding the algebraic curve is very straightforward. We make an ansatz $H(\UU,\VV;Q)=\sum_{i=0}^{d_\UU} \sum_{j=0}^{d_\VV} c_{ij} \UU^i \VV^j$ and solve for the ($Q$-dependent) coefficients $c_{ij}$ in a series expansion of $H(\UU,\VV(\UU);Q)$. Of course the fact that $d_\UU,d_\VV$ are finite is non-trivial. For the case at hand, one can show that for $(r,s)$ being coprime the degrees are given by
\eqn\degs{d_\beta=\left(s+r \atop r\right)=\left(s+r \atop s\right),\qquad d_\alpha={d_\beta \over r+s}\ ,}
while if $(r,s)$ are not coprime there is a simple bound $d_\beta(r,s)\le {\rm Min}(d_\beta(r,s+1),d_\beta(r+1,s))$ while $d_\alpha={d_\beta\over s+r}$ still holds.

As an example, we have extracted the mirror curve associated with the torus knot $\CK_{3,4}$, purely based on the knowledge of the genus zero disk instantons coming from \bva. For this case we find the following mirror curve
\eqn\Fthrfour{\eqalign{H_{3,4}(\UU,\VV;Q)=&\big(1 - Q \VV\big)  -\VV^{4}\big(1-\VV+\VV^{2} - 2 Q \VV^2 - 5 \VV^3 + 9 Q \VV^3 - 3 Q^2 \VV^3  \cr
&  + 4 Q \VV^4 - 4 Q^2 \VV^4 - Q^2 \VV^5 +
 Q^3 \VV^5 - 3 Q^3 \VV^6 + 3 Q^4 \VV^6+Q^5 \VV^8 \cr
& - Q^6 \VV^9\big)\UU-\VV^{10}\big(1 - \VV - 3 \VV^2 + 4 Q \VV^2 + 4 \VV^3 - 8 Q \VV^3 + 3 Q^2 \VV^3 \cr
& - 10 \VV^4 + 27 Q \VV^4 - 17 Q^2 \VV^4 + 6 Q \VV^5 - 10 Q^2 \VV^5 +
 4 Q^3 \VV^5  \cr
& - 3 Q^2 \VV^6 + 5 Q^3 \VV^6 - 3 Q^4 \VV^6 - 5 Q^3 \VV^7 + 6 Q^4 \VV^7 + Q^4 \VV^8  \cr
&-  Q^5 \VV^9\big)\UU^{2} +\VV^{16}\big(1- \VV + 5 \VV^2 - 6 Q \VV^2 + 3 \VV^3 - 5 Q \VV^3 + 3 Q^2 \VV^3  \cr
&- 6 \VV^4 + 10 Q \VV^4 - 4 Q^2 \VV^4 + 10 \VV^5 - 27 Q \VV^5 + 17 Q^2 \VV^5 - 4 Q \VV^6  \cr
&+ 8 Q^2 \VV^6 - 3 Q^3 \VV^6 + 3 Q^2 \VV^7 - 4 Q^3 \VV^7 + Q^3 \VV^8 - Q^4 \VV^9\big)\UU^{3}+\VV^{22} \cr
&\big(1- \VV + 3 \VV^3 - 3 Q \VV^3 + \VV^4 - Q \VV^4 - 4 \VV^5 + 4 Q \VV^5 + 5 \VV^6 - 9 Q \VV^6  \cr
&+ 3 Q^2 \VV^6 - Q \VV^7  + 2 Q^2 \VV^7+ Q^2 \VV^8 - Q^3 \VV^9\big)\UU^{4}-\VV^{34}\big(1-\VV\big)\UU^{5}\ .}}
At $Q=1$, the above curve factorizes to several irreducible components and it contains a factor of the full A-polynomial of the torus knot $\CK_{3,4}$
\eqn\FthrfourQ{H_{3,4}(\UU,\VV;1)=(1 - \VV) (1 - \UU \VV^4) (1 - \UU \VV^6) (1 + \UU \VV^6)^2 (1 - \UU \VV^{12})\ .}
The mirror curve associated to torus links can also be constructed from the genus zero disk instantons. The simplest torus link is the link $\CK_{2,2}$, {\it i.e.}, the Hopf link. For this link, we find the following curve
\eqn\Hopfcur{\eqalign{H_{2,2}(\UU,\VV;Q)=&\big(1-Q\VV\big)\big(1+Q\VV\big)+\VV^{2}
\big(1 + 2 \VV^2 - 4 Q \VV^2 + Q^2 \VV^4\big)\UU\cr
&-\VV^{6}\big(1-\VV\big)\big(1+\VV\big)\UU^{2}\ .}}
This curve degenerates at $Q=1$ to the following components
\eqn\HopfQ{H_{2,2}(\UU,\VV;1)=(1 - \VV) (1 + \VV) (1 + \UU \VV^2) (1 -\UU \VV^4)\ .}
As suggested in \AVtwo, a possible interpretation for the leading factors $(1-\VV)$ and $(1+\VV)$ of \HopfQ\ could be the identification with branches of the moduli space, where a component of the A-model Lagrangian cycle leaves the zero section $S^{3}$ of the deformed conifold. Since the Hopf link is constructed out of two A-model Lagrangian components, it is suggestive that the individual factors $(1-\VV)$ and $(1+\VV)$ are associated to the two A-model Lagrangian components, respectively. This claim is further supported by the observation that the disk amplitudes $\log \VV_\pm(\UU)$ extracted from the vicinities $\VV_\pm(0)=\pm Q$ of the spectral curve $H_{2,2}(\UU,\VV_\pm(\UU);Q)=0$ obey $\VV_+(\UU) = - \VV_-(\UU)$.

We have computed the spectral curves associated to other torus knots from the disk amplitude. In particular the entire spectral curve $H_{3,5}(\UU,\VV;Q)=0$ of the torus knot $\CK_{3,5}$ is recorded in full glory in appendix~B. As can be seen from the general degree \degs, for higher integers $(r,s)$ these curves become too lengthy to be spelled out here explicitly. Therefore, we just present their degeneration at $Q=1$ here
\eqn\TotuscurveQ{\eqalign{
&H_{2,9}(\UU,\VV;1)=(1 - \VV) (1 + \UU \VV^9)^4 (1 - \UU \VV^{18})\ ,\cr
&H_{2,11}(\UU,\VV;1)=(1 - \VV) (1 + \UU \VV^{11})^5 (1 - \UU \VV^{22})\ ,\cr
&H_{2,13}(\UU,\VV;1)=(1 - \VV) (1 + \UU \VV^{13})^6 (1 - \UU \VV^{26}) \ ,\cr
&H_{2,15}(\UU,\VV;1)=(1 - \VV) (1 + \UU \VV^{15})^7 (1 - \UU \VV^{30})\ ,\cr
&H_{3,5}(\UU,\VV;1)=(1 - \VV) (1 - \UU \VV^5)^2 (1 - \UU \VV^{15}) (1 - \UU^2 \VV^{15})^2\ ,\cr
&H_{3,7}(\UU,\VV;1)=(1 - \VV) (1 - \UU \VV^7)^5 (1 - \UU \VV^{21}) (1 - \UU^2 \VV^{21})^3\ ,\cr
&H_{3,3}(\UU,\VV;1)=(1 - \VV) (1 + \VV + \VV^2) (1 - \UU \VV^3) (1 - \UU \VV^9) (1 - \UU^2 \VV^9)\ ,\cr
&H_{3,6}(\UU,\VV;1)=(1 - \VV) (1 + \VV + \VV^2) (1 - \UU \VV^6)^4 (1 - \UU \VV^9)^2 (1 + \UU \VV^9)^2 (1 - \UU \VV^{18})\ .}}
Note that all these curves contain (up to a framing transformation) a factor of the A-polynomial of the their corresponding torus knots/links, and they exhibit the formula \degs\ for their degrees.

The calculated defining polynomials of the spectral curves $H_{r,s}(\UU,\VV)$ agree with the known augmentation polynomials, which are directly defined in terms of knot contact homology for knots $\CK$ on $S^3$ in \refs{\Ng}. This relationship between augmentation polynomials and the B-model spectral curves has been conjectured in \refs{\AVtwo,\NgJX} and has been studied for torus knots $\CK_{2,2p+1}$ (with integral $p$) in \FujiNX.

Our calculation demonstrates the general fact that the spectral curve --- or at least a relevant irreducible component --- can be reconstructed uniquely from the knowledge of the entire disk amplitude. Choosing different expansion points corresponds to different phases in the open string moduli space and leads to different integer open string invariants.

The next step is to reconstruct the quantum information from the spectral curve using the Eynard-Oratin method. In particular the annulus amplitude is the first non-trivial piece of quantum information, which goes beyond the data encoded in the HOMFLY polynomials that are merely colored with symmetric representations. In particular, it is possible to apply the Eynard-Oratin formalism to the spectral curve $H_{2,3}(\UU,\VV;Q)=0$ associated to the torus knot $\CK_{2,3}$, as this spectral curve is of geometric genus one \WProg.

\newsec{Torus Knots in Lens Spaces}

The main focus of the previous sections was to consider torus knots in $S^{3}$, and for this purpose, we had dealt with the resolved conifold geometry. However, the construction of the Lagrangian cycle of section 2 can be generalized to other toric geometries as well. An interesting example to consider is knots/links in lens spaces. It turns out that lens spaces can be constructed as a torus bundle over an interval, very similar to the construction of section 2. Let us briefly mention how the geometric setup will work. As we saw in section 2, $S^{3}$ can be represented as a torus bundle over an interval so that the $(1,0)$ cycle of the torus shrinks at one end point of the interval and the $(0,1)$ cycle shrinks at the other end. Now, instead of degenerating $(0,1)$ cycle, we pick a different cycle, namely the $(p,q)$ cycle of the torus. This leads to the lens space $L(q,p)$ whose fundamental group is $\pi_{1}(L(q,p))=\IZ_{q}$. To explain why this is the case, let us cut the interval (base of the torus fibration) into two pieces. Upon this operation, each half is a $T^{2}=S^{1}\times S^{1}_{c}$ fibration over an interval where $S^{1}_{c}$ is a contractible one cycle of the torus. Therefore, each half has the topology of a solid torus. In the case of $S^{3}$, the contractible and the non-contractible one-cycles of the left and the right solid tori are exchanged. This corresponds to performing an $S$ operation in gluing the two solid tori and this exactly describes the genus one Heegaard splitting of $S^{3}$. Now, instead of degenerating $(0,1)$ cycle of the right solid torus, if we degenerate the $(p,q)$ cycle, we would need to perform a general $SL(2,\IZ)$ operation in gluing the two solid tori. This exactly corresponds to the Heegaard splitting of lens spaces. Once we have constructed the lens space $L(q,p)$, we can embed this geometry into the A-model topological strings with the target space $T^{*}L(q,p)$. After the large $N$ transition, we end up with a new geometry, and we obtain the invariants of torus knots by computing the open-string amplitudes of the corresponding resolved geometry, as in section~\secVertex.

The simplest lens space to consider is $L(2,1)\cong \IR\IP^{3}$. This lens space can be obtained by the antipodal identification of the points on a three-sphere $L(2,1)=S^{3}/{\IZ_{2}}$. To fit this in our setup, we can view this space as a torus bundle over a segment, where $(1,0)$ cycle of the torus fiber shrinks at one end, and the $(1,2)$ cycle at the other end. Let first consider the unknot in this lens space. This case has first been studied in detail \AKMVtwo. Chern-Simons theory on this space is alternatively described by an open A-model topological string theory with the target space $T^{*}(S^{3}/\IZ_{2})$. Upon the large $N$ transition, the local $\IP^{1}\times\IP^{1}$ geometry emerges, and the toric Lagrangian brane which lands on one of the outer legs of this geometry is supposed to capture the invariants of the framed unknot in this lens space \BKMP. We can now apply our method to construct the toric Lagrangian brane whose open-string amplitudes produce the invariants of torus knots in the lens space $L(2,1)$\foot{It would also be desirable to construct the analog of the Lagrangian of \DSV~for this lens space.}. We will do this in two different ways. We first construct the Lagrangian cycle describing torus knots in $L(2,1)$, using the explicit construction of section 2, and then we check this against the Lagrangian cycle one finds by the B-model argument of \BEM~and mirror symmetry.

\subsec{Construction of the Lagrangian Cycle in $L(2,1)$}

As mentioned earlier, the bulk geometry before the large $N$ transition can be thought as the total space of $T^{*}L(2,1)$. As mentioned above, we represent the base $L(2,1)$ of the fibration as a torus bundle over an interval, where $(1,0)$ and $(1,2)$ cycles are shrunken, one at each end point. To realize an $(r,s)$ torus knot in this lens space, we choose the cycle $r\,\ac+s\,\bc$ to be away from the fixed points of the $\IZ_{2}$ action, and then the rest of the story of section 2 would go in the same way. We will eventually end up with a Lagrangian describing an $(r,s)$ torus knot in $L(2,1)$. Now, let us focus on the open/closed geometry after the large $N$ transition.

The bulk geometry in this case is the total space of the fibration $\CO(-2,-2)\rightarrow\IP^{1}\times\IP^{1}$ which is described in terms of the two following GLSM charge vectors $P^{1}=(-2,1,1,0,0)$ and $P^{2}=(-2,0,0,1,1)$
\eqn\Fbulk{\eqalign{&|X_{1}|^{2}+|X_{2}|^{2}-2|X_{0}|^{2}={\rm Re}\,t_{1}\ ,\cr
&|X_{3}|^{2}+|X_{4}|^{2}-2|X_{0}|^{2}={\rm Re}\,t_{2}\ ,}}
where $t_{1}$ and $t_{2}$ are the two K\"{a}hler moduli of the bulk geometry, and in order to make contact with Chern-Simons theory on $S^{3}/\IZ_{2}$, one has to set $t_{1}=-t_{2}$. To define the Lagrangian submanifold in this geometry, we introduce with the following Lagrangian charges $p^{1}=(-1,1,0,0,0)$ and $p^{2}=(-1,0,0,1,0)$
\eqn\Flag{\eqalign{&|X_{1}|^{2}-|X_{0}|^{2}={\rm Re}\,c_{1}\ ,\cr
&|X_{3}|^{2}-|X_{0}|^{2}={\rm Re}\,c_{2}\ ,}}
where ${\rm Re}\,c_{1}\geq0$ and ${\rm Re}\,c_{2}\geq0$. The above Lagrangian cycle degenerates and will have the topology $S^{1}\times\IR^{2}$, when we set $c_{1}=c_{2}$. In this limit, the Lagrangian, which describes the unknot in $S^{3}/{\IZ_{2}}$, lands on one of the outer legs of the toric skeleton (see \lfig\ZnuPP).

To study the torus knots, we are supposed to perform an $SL(2,\IZ)$ on the Lagrangian charges $\pmatrix{p^{1} \cr p^{2}}\mapsto \pmatrix{\tilde{p}^{1} \cr \tilde{p}^{2}}=M\pmatrix{p^{1} \cr p^{2}}$ in which $M$ is given by \SLtwoZ. This results in $\tilde{p}^{1}=(-r-s,r,0,s,0)$ and $\tilde{p}^{2}=(-p-q,p,0,q,0)$
\eqn\Flach{\eqalign{&r|X_{1}|^{2}+s|X_{3}|^{2}-(r+s)|X_{0}|^{2}={\rm Re}\,\tilde{c}_{1}\ ,\cr &p|X_{1}|^{2}+q|X_{3}|^{2}-(p+q)|X_{0}|^{2}={\rm Re}\,\tilde{c}_{2}\ .}}
As in section 2, we can undo the $SL(2,\IZ)$ transformation on the Lagrangian charges and instead impose the transformation on the open moduli space. In this passive form, one finds the old Lagrangian charges $p^{1}$ and $p^{2}$ with the following open moduli
\eqn\Fmod{\pmatrix{\tilde{c}_{1} \cr \tilde{c}_{2}}\mapsto \pmatrix{\hat{c}_{1} \cr \hat{c}_{2}}=M^{-1}\pmatrix{\tilde{c}_{1} \cr \tilde{c}_{2}}\ .}
To get the topology $S^{1}\times\IR^{2}$, we degenerate the Lagrangian by setting $\hat{c}_{1}=\hat{c}_{2}$. This results in $\hat{c}_{1}=\hat{c}_{2}={1\over r+s}\tilde{c}_{1}$. We notice that an $(r,s)$ torus knot does not have a unique representation as an $SL(2,\IZ)$ matrix. If $M$ represents an $(r,s)$ torus knot, then $MT^{m}$ represents the same knot but with different units of framing. Therefore, we choose the representative $MT^{-1}$, and this leads us to $\hat{c}_{1}={1\over s}\tilde{c}_{1}$. As we will see, the same manipulation occurs when we construct the Lagrangian cycle by means of mirror symmetry.

\subsec{Lagrangian Cycle in $L(2,1)$ from Mirror Symmetry}

The B-model derivation of \BEM~for the mirror curve that describes a torus knot is in fact quite general and is not specific to the resolved conifold geometry. Of course, not all toric examples have a knot theory interpretation. But one can definitely apply the B-model argument of \BEM~for the case of local $\IP^{1}\times\IP^{1}$ in order to describe torus knots in the lens space $L(2,1)$. According to the philosophy of \BEM, to make contact with torus knots, we have to make an $SL(2,\IZ)$ transformation on the local coordinates of the mirror B-model. For definiteness, we assume that we are working in the patch $y_{0}\neq 0$, and similar to the previous case, we define the corresponding affine coordinates
\eqn\RPaff{\alpha={y_{1}\over y_{0}}\quad,\quad \gamma={y_{2}\over y_{0}}\quad,\quad \beta={y_{3}\over y_{0}}\quad,\quad \lambda={y_{4}\over y_{0}}\ .}
In this local patch of coordinates, we perform the transformation
\eqn\Ftrans{\pmatrix{\alpha \cr \beta}\mapsto \pmatrix{\hat{\alpha} \cr \hat{\beta}}=\pmatrix{\alpha^{r}\,\beta^{s}\cr \alpha^{p}\,\beta^{q}}\ .}
As in the case of the resolved conifold, this transformation should presrve the bulk geometry. To observe this, we first perform a general transformation on $\gamma$ and $\lambda$ in this patch
\eqn\Ylef{\eqalign{\gamma&=\hat{\alpha}^{a}\,\hat{\gamma}^{b}\,\hat{\beta}^{c}\,
\hat{\lambda}^{d}\ ,\cr \lambda&=\hat{\alpha}^{e}\,\hat{\gamma}^{f}\,\hat{\beta}^{g}\,\hat{\lambda}^{h}\ ,}}
where $a$, $b$, $c$, $d$, $e$, $f$, $g$, and $h$ are some integers. Now, in order to read off the A-model charge vectors, we use the first equation of \Qcons~
in terms of the homogeneous coordinates. Doing that, we obtain
\eqn\FPtrans{\eqalign{&\hat{y}_{0}^{-q+s-a-b-c-d}\,\hat{y}_{1}^{q+a}\,\hat{y}_{2}^{b}
\,\hat{y}_{3}^{c-s}\,\hat{y}_{4}^{d}=Q_{1}\ ,\cr
&\hat{y}_{0}^{p-r-e-f-g-h}\,\hat{y}_{1}^{-p+e}\,\hat{y}_{2}^{f}\,
\hat{y}_{3}^{r+g}\,\hat{y}_{4}^{h}=Q_{2}\ ,}}
where $Q_{1}=e^{-t_{1}}$ and $Q_{2}=e^{-t_{2}}$. To preserve the bulk geometry, we must require the new charge vectors to be the same as $P^{1}=(-2,1,1,0,0)$ and $P^{2}=(-2,0,0,1,1)$. This uniquely determines the above unknown integers and we gain
\eqn\Fabcd{\eqalign{& a=1-q\qquad,\qquad b=1\qquad,\qquad c=s\qquad,\qquad d=0\ ,\cr
& e=p\qquad,\qquad f=0\qquad,\qquad g=1-r\qquad,\qquad h=1\ .}}
Using \SLtwoZ, we can express $\hat{Y}_{2}$ and $\hat{Y}_{4}$ coordinates in terms of the old coordinates
\eqn\FYold{\hat{\gamma}=\alpha^{1-r}\,\gamma\,\beta^{-s}\qquad,\qquad \hat{\lambda}=\alpha^{-p}\,\beta^{1-q}\,\lambda\ .}

We now proceed with the analysis of the open-string sector. As in the previous example, the mirror B-brane is constructed using the second equation of \Qcons.
Let us work in the same local patch. In this patch, the Lagrangian charges $p^{1}$ and $p^{2}$ give us the following constraints
\eqn\Ftranlag{\alpha=e^{c_{1}}\qquad,\qquad \beta=e^{c_{2}}\ .}
We are now supposed to perform the same $SL(2,\IZ)$ transformation on $Y_{i}$ coordinates for the open-string sector. Using transformation \Ftrans~on the $Y_{i}$ coordinates and rewriting the result in terms of the original homogeneous coordinates $y_{i}$, we obtain
\eqn\Fnewlagchar{\eqalign{&\hat{y}_{0}^{s-q}\,\hat{y}_{1}^{q}\,\hat{y}_{3}^{-s}=e^{c_{1}}\ ,\cr &\hat{y}_{0}^{p-r}\,\hat{y}_{1}^{-p}\,\hat{y}_{3}^{r}=e^{c_{2}}\ .}}
From \Fnewlagchar, we can read off the new Lagrangian charge vectors corresponding to the transformed Lagrangian. The new Lagrangian is described by two charge vectors $\tilde{p}^{1}=(s-q,q,0,-s,0)$ and $\tilde{p}^{2}=(p-r,-p,0,r,0)$, and similar to the previous case, this is a special Lagrangian cycle
\eqn\Flagchar{\eqalign{&(s-q)|\hat{X}_{0}|^{2}+q|\hat{X}_{1}|^{2}-s|\hat{X}_{3}|^{2}
={\rm Re}\,\tilde{c}_{1}\ ,\cr
& (p-r)|\hat{X}_{0}|^{2}-p|\hat{X}_{1}|^{2}+r|\hat{X}_{3}|^{2}={\rm Re}\,\tilde{c}_{2}\ .}}
To analyze the topology of the Lagrangian, we again use the same trick to undo the $SL(2,\IZ)$ transformation. Similar to the previous case, instead of working with   $\tilde{p}^{1}$ and $\tilde{p}^{2}$, we can equivalently work with the charge vectors $r\tilde{p}^{1}+s\tilde{p}^{2}$ and $p\tilde{p}^{1}+q\tilde{p}^{2}$. This amounts to performing the following transformation in the moduli space of the Lagrangian cycle
\eqn\Fctran{\pmatrix{c_{1} \cr c_{2}}\mapsto \pmatrix{\hat{c}_{1} \cr \hat{c}_{2}}=M\pmatrix{\tilde{c}_{1} \cr \tilde{c}_{2}}\ .}
As before, for generic values of $\tilde{c}_{1}$ and $\tilde{c}_{2}$, the Lagrangian will have the topology $T^{2}\times\IR$. However, when it lands on an outer leg, the topology will be $S^{1}\times\IR^{2}$, as required. This implies that we should be working in the phase where $\hat{c}_{1}=\hat{c}_{2}$. Using \Fctran, we conclude that $\hat{c}_{1}=\hat{c}_{2}={1\over q-s}\tilde{c}_{1}$. Here, as in the A-model analysis of the previous section, we change the representation of the torus knot from $M$ to $T^{-1}M$. This would then imply that $\hat{c}_{1}=-{1\over s}\tilde{c}_{1}$ (the minus sign only reverses the orientation of the knot). As in the case of the conifold, this rescaling would lead to appearance of a fractional framing for the knot (the argument of section 2.2 will be still valid so long as we keep the torus knot away from the fixed points of the $\IZ_{2}$ action).

Therefore, the whole argument of section 3 for the computation of the open-string amplitude would go through and one will find the following structure for the open-string partition function
\eqn\Fdec{Z=\sum_{\mu}\sum_{|\nu|=s|\mu|}c_{\mu,s}^{\nu}\,Z_{\nu}{\rm Tr}_{\mu}{\cal V}\ ,}
where $Z_{\nu}$ is the open-string amplitude corresponding to the diagram depicted in \lfig\ZnuPP, where the outer brane has the framing $r/s$.

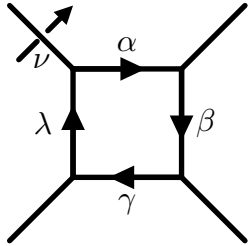
\figboxinsert\ZnuPP{The open-string amplitude $Z_{\nu}$ in local $\IP^{1}\times\IP^{1}$.}{
\DrawDiag{\drawdim cm \linewd 0.06
\move(2 3.4)
\move(2 2) \lvec(3.2 2) \rlvec(0 -1.2) \rlvec(-1.2 0) \rlvec(0 1.2) \rlvec(-0.7 0.7) \move(3.2 2) \rlvec(0.7 0.7) \move(3.2 0.8) \rlvec(0.7 -0.7) \move(2 0.8)
\rlvec(-0.7 -0.7) \move(1.4 2.1) \rlvec(0.2 0.2) \move(1.72 2.42)
\arrowheadsize l:0.2 w:0.2 \arrowheadtype t:F \avec(2 2.7)
\move(2 2) \arrowheadsize l:0.28 w:0.26 \avec(2.8 2)
\move(3.2 2) \avec(3.2 1.2) \move(3.2 0.8) \avec(2.4 0.8) \move(2 0.8) \avec(2 1.6)
\textref h:C v:C \htext(1.65 2.08){$\nu$}
\textref h:C v:C \htext(2.6 2.28){$\alpha$}
\textref h:C v:C \htext(3.47 1.4){$\beta$}
\textref h:C v:C \htext(2.6 0.5){$\gamma$}
\textref h:C v:C \htext(1.67 1.4){$\lambda$}
\move(0 0)
}}

It was shown in \AKMVtwo\ that Chern-Simons theory defined on the Lens space $S^{3}/{\IZ_{2}}$ agrees with the open string amplitudes of the above geometry in the {\it orbifold phase} of the closed-string moduli space. This implies that, as far as knot invariants are concerned, what topological vertex computes for this geometry is not going to be useful, as topological vertex computes the amplitudes in the large radius limit. It would be very interesting to see if one can compute this amplitude with the techniques developed in the orbifold topological vertex \BCY, and reproduce the HOMFLY polynomial of torus knots in this lens space.\foot{As shown in \BKMP, one can carry out the B-model computations in any corner of the moduli space. Therefore, using the topological recursion of \EO\ would be  another alternative.} It has also been shown in \refs{\Br,\BC}~that for toric open/closed geometries, one can perform an equivariant localization computation in the orbifold phases of the open/closed moduli space. As an another alternative, one can employ this approach to learn about knot invariants in lens spaces \BCprog.

\newsec{Concluding Remarks}

In this note, we have shown how the (colored) HOMFLY polynomial of a given torus knot is captured via the topological vertex, directly in the A-model picture. The key idea has been the implementation of the B-model $SL(2,\IZ)$ transformation of \BEM\ into the A-model. Using local mirror symmetry, we have demonstrated that the constructed toric A-branes associated to torus knots are mirror-symmetric to the B-model spectral curve of \BEM. The method we have developed is naturally generalized to other toric geometries, and this gives a handle to describe torus knots in certain three-manifolds such as lens spaces.

There are many open questions to be pursued in the future. It would be interesting to further develop the correspondence between knots on $S^3$ and their associated spectral curves as proposed in \AVtwo\ and as further analyzed here in section~4. In particular, it would be desirable to understand systematically the structure of the B-model spectral curves associated to knots on $S^3$. We have proposed in section~4.3 a formula, which predicts the degrees of the spectral curves associated to torus knots $\CK_{r,s}$ according to the integers $r,s$. But in order to arrive at the explicit spectral curve, it is still necessary to solve a system of linear equations with many unknowns, which becomes a computational extensive procedure for knot $\CK_{r,s}$ with large integers $r,s$. Presumably, progress in this direction can be achieved by better understanding all the components of the B-model spectral curves that emerge in the conifold degeneration limit. Clearly, it would be very interesting to derive spectral curves associated to knots, which do not fall in the class of torus knots. In \AVtwo\ a spectral curve was derived for the figure eight knot on $S^3$ because its HOMFLY polynomials colored with symmetric representations are known. However, the benefit of the B-model spectral curves would really unfold, if one directly deduces the spectral curve from the knot on $S^3$. Then the spectral curve of such knots could be employed to predict HOMFLY polynomials colored with arbitrary representations. The discussed conjectured relationship between augmentation polynomials of knots on $S^3$ and their B-model spectral curves promises the realization of such an approach \refs{\AVtwo,\NgJX}.

Another obvious line for future research directions concerns various aspects of the refinement. The relation between the HOMFLY polynomial associated to algebraic links of plane curve singularities and framed stable pair invariants for a reduced curve on the blown up conifold was recently proven in \Maul. The refinement of the symmetric obstruction theory of stable pairs by the virtual Bialynicki-Birula decomposition \CKK\ should give a systematic way to obtain the refined invariants.

Furthermore, it turns out that the refined topological vertex \IKV\ computes the Khovanov-Rozansky invariant of the unknot (and the Hopf link) colored by an arbitrary representation \GIKV. In a separate line of developments, it has recently been observed that the superpolynomial of torus knots has indeed a decomposition in terms of superpolynomials of colored unknots \GAF. The decomposition, however, involves a new ingredient, namely the appearance of gamma-factors \GAF.\foot{The gamma-factors for any torus knot $\CK_{2,2p+1}$ (colored by totally symmetric or totally antisymmetric representations) have recently been computed in closed forms in \FGS.} Although one is able to compute the gamma-factors in a recursive manner, one cannot derive them from first principles. On the other hand, one can compute the refined open-string amplitudes associated with the toric Lagrangian \newlagchar, using the refined topological vertex, and in particular, considering the observation made in \IK.\foot{It has been shown in \IK~that for the case of unknot and the Hopf link, the refined normalized open-string amplitudes of the resolved conifold geometry reproduce the S-matrix elements of the refined Chern-Simons theory \AS.} The result is expected to reproduce the colored superpolynomial of torus knots. The only subtlety in this computation is the framing factor of the Lagrangian \newlagchar. Recall that in the ordinary case, the right choice of framing relies on the gauge theory formulation of Chern-Simons theory (see section 3.2). However, in the refined case, the framing transformation is not very well understood and furthermore, we are lacking a gauge theory formulation of the refined Chern-Simons theory \AS. Nonetheless, if one succeeds to compute the refined open-string amplitudes of the Lagrangian \newlagchar~with the right framing property, one may hope to provide a better explanation for the appearance of the gamma-factors.

It is well known that the observables of Chern-Simons gauge theory with orthogonal gauge groups lead to Kauffman polynomial of knots and links. For torus knots, it has been shown in \refs{\Stev, \ChCh} that the Kauffman polynomial of a given torus knot/link has a decomposition in terms of the quantum dimensions of the orthogonal group. On the other hand, the real topological vertex \KPW, computes the topological amplitudes of toric Calabi-Yau geometries in the presence of D-branes and O-planes. Applying the real vertex, it would be very interesting to derive the analog of the Rosso and Jones formula for the Kauffman polynomial of torus knots using the new toric Lagrangian \newlagchar.

\vskip 1cm
\noindent
{\bf Acknowledgements}
\vskip 0.2cm

We would like to thank
Jim Bryan, Christoph Koutschan, Marcos Mari\~no, and
Vivek Shende for discussions and correspondence. This work is supported by the DFG grant KL2271/1-1.
\bigskip

\appendix{A}{Adams Operation}

In this appendix, we provide more details on the coefficients of Adams operation. Let $G$ be a group and ${\rm ch}_{\pi}$ be a character of $G$ associated with the representation $\pi$ of $G$. In the context of group theory, it is shown that for any nonnegative integer $k$, $g\mapsto{\rm ch}_{\pi}^{k}(g)\equiv{\rm ch}_{\pi}(g^{k})$ is a virtual character of $G$.

We would now like to specialize to the case where ${\rm ch_{\pi}}$ is the character of a representation $\pi: G\rightarrow GL(n,\IC)$. Let us consider the ring of symmetric polynomials in $n$ variables with integer coefficients and let $f(x_{1},\cdots,x_{n})$ be an element of this ring. It is then shown that $\psi_{f}(g)=f(t_{1},\cdots,t_{n})$ where $g\in GL(n,\IC)$ and $t_{1},\cdots,t_{n}$ are eigenvalues of $g$ is a virtual character of $GL(n,\IC)$. An operation ${\rm ch}_{\pi}\mapsto{\rm ch}_{\pi}^{k}$ on the ring of virtual characters of $G$ is referred as an Adams operation. This is a ring homomorphism and upon choosing a basis for the ring of virtual characters, one has the following decomposition
\eqn\ringdec{{\rm ch}_{\pi}^{k}=\sum_{\sigma}\alpha_{\pi,k}^{\sigma}\,{\rm ch}_{\sigma}\ ,}
where $\alpha_{\pi,k}^{\sigma}$ are the coefficients of Adams operation. These coefficients can be conveniently computed. For the case of the unitary subgroup of $GL(n,\IC)$, these coefficients can be easily expressed in terms of the characters of the symmetric group. Having in mind the orthogonality relations of the first and second kinds for the characters of the symmetric group
\eqn\orthone{\sum_{\vec{k}}{1\over z_{\vec{k}}}\,\chi_{\mu}(C_{\vec{k}})\chi_{\nu}(C_{\vec{k}})
=\delta_{\mu,\nu}\ ,}
and
\eqn\orthtwo{\sum_{\nu}\chi_{\nu}(C_{\vec{k}_{1}})\chi_{\nu}(C_{\vec{k}_{2}})=
z_{\vec{k}_{1}}\delta_{\vec{k}_{1},\vec{k}_{2}}\ ,}
one can easily compute the coefficients of Adams operation for the unitary groups by means of the Frobenius formula \Stev
\eqn\adamcoeff{c_{\mu,n}^{\nu}=\sum_{|\alpha|=|\mu|}{1\over z_{\vec{k}_{\alpha}}}\chi_{\mu}(C_{{\vec{k}}_{\alpha}})
\chi_{\nu}(C_{{\vec{k}}_{n\alpha}})\  .}
These coefficients have several interesting properties. First of all, $c_{\mu,1}^{\nu}=\delta^{\nu}_{\mu}$ as expected. Second, $c_{\mu,n}^{\nu}$ is nonzero only if $|\nu|=n|\mu|$. In addition, it turns out that these coefficients enjoy the following property $\sum_{\nu}c_{\mu,n}^{\nu}\,c_{\nu,m}^{\lambda}=c_{\mu,nm}^{\lambda}$
\eqn\adamsq{\eqalign{\sum_{\nu}c_{\mu,n}^{\nu}\,c_{\nu,m}^{\lambda}&=\sum_{\nu}
\Big(\sum_{|\alpha|=|\mu|}{1\over z_{\vec{k}_{\alpha}}}\chi_{\mu}(C_{\vec{k}_{\alpha}})\chi_{\nu}(C_{\vec{k}_{n\alpha}})
\Big)\Big(\sum_{|\beta|
=|\nu|}{1\over z_{\vec{k}_{\beta}}}\chi_{\nu}(C_{\vec{k}_{\beta}})\chi_{\lambda}(C_{\vec{k}_{m\beta}})
\Big)\cr
&=\sum_{|\alpha|=|\mu|}\sum_{|\beta|=n|\mu|}{1\over z_{\vec{k}_{\alpha}}z_{\vec{k}_{\beta}}}\chi_{\mu}(C_{\vec{k}_{\alpha}})
\chi_{\lambda}(C_{\vec{k}_{m\beta}})\sum_{\nu}\chi_{\nu}(C_{\vec{k}_{n\alpha}})
\chi_{\nu}(C_{\vec{k}_{\beta}}) \cr
&=\sum_{|\alpha|=|\mu|}{1\over z_{\vec{k}_{\alpha}}}\chi_{\mu}(C_{\vec{k}_{\alpha}}) \chi_{\lambda}(C_{\vec{k}_{nm\alpha}})=c_{\mu,nm}^{\lambda}\ .}}

\vfill\break
\appendix{B}{Spectral Curve of the Torus Knot $\CK_{3,5}$}
\noindent
The spectral curve $H_{3,5}$ associated to the torus knot $\CK_{3,5}$ reads:
\eqn\Fthrfive{\eqalign{H_{3,5}(\UU,\VV;Q)=&\big(1 - Q \VV\big)-\VV^{5}\big(2 - \VV - Q \VV + \VV^2 - 2 Q \VV^2 + Q^2 \VV^2 -
 7 \VV^3 + 14 Q \VV^3 \cr
& - 7 Q^2 \VV^3 + 6 Q \VV^4 -
 9 Q^2 \VV^4 + 3 Q^3 \VV^4 - Q^2 \VV^5 +
 4 Q^3 \VV^5 - 3 Q^4 \VV^5 \cr
& - 5 Q^3 \VV^6 +
 5 Q^4 \VV^6 + 3 Q^4 \VV^7 - 3 Q^5 \VV^7 -
 3 Q^5 \VV^8 + 3 Q^6 \VV^8 + Q^7 \VV^{10}\cr
& - Q^8 \VV^{11}\big)\UU+\VV^{10}\big(1 - \VV - 3 \VV^3 + 6 Q \VV^3 - 3 Q^2 \VV^3 + 5 \VV^4 - 4 Q \VV^4 \cr
&- Q^2 \VV^4 - 6 \VV^5 +
 17 Q \VV^5 - 13 Q^2 \VV^5 + 21 \VV^6 - 70 Q \VV^6 + 69 Q^2 \VV^6 \cr
& - 21 Q^3 \VV^6 + 3 Q^4 \VV^6 -
 15 Q \VV^7 + 38 Q^2 \VV^7 - 25 Q^3 \VV^7 + 2 Q^4 \VV^7 \cr
& + 5 Q^2 \VV^8 - 19 Q^3 \VV^8 + 14 Q^4 \VV^8 + 18 Q^3 \VV^9 - 34 Q^4 \VV^9 + 17 Q^5 \VV^9 \cr
& - Q^6 \VV^9 - 9 Q^4 \VV^{10} + 11 Q^5 \VV^{10} + 5 Q^5 \VV^{11} - 7 Q^6 \VV^{11} + 3 Q^6 \VV^{12} \cr
& - 3 Q^7 \VV^{12}\big)\UU^{2} -\VV^{19}\big(3 - 3 Q + 3 \VV - 7 Q \VV - 8 \VV^2 + 12 Q \VV^2 - 10 \VV^3 \cr
& + 22 Q \VV^3 - 11 Q^2 \VV^3 - Q^3 \VV^3 +
 15 \VV^4 - 48 Q \VV^4 + 42 Q^2 \VV^4 - 9 Q^3 \VV^4  \cr
& -35 \VV^5 + 140 Q \VV^5 - 172 Q^2 \VV^5 + 67 Q^3 \VV^5 + 20 Q \VV^6 - 62 Q^2 \VV^6 \cr
& + 52 Q^3 \VV^6 - 12 Q^4 \VV^6 + 3 Q^5 \VV^6 - 10 Q^2 \VV^7 + 32 Q^3 \VV^7 - 23 Q^4 \VV^7 \cr
& - 22 Q^3 \VV^8 + 47 Q^4 \VV^8 - 25 Q^5 \VV^8 + 9 Q^4 \VV^9 - 15 Q^5 \VV^9 + 6 Q^6 \VV^9 \cr
& - 2 Q^6 \VV^{11} + 2 Q^7 \VV^{12}\big)\UU^{3} -\VV^{25}\big(2 - 2 \VV + 9 \VV^3 - 15 Q \VV^3 + 6 Q^2 \VV^3 \cr
& - 22 \VV^4 + 47 Q \VV^4 - 25 Q^2 \VV^4 - 10 \VV^5 + 32 Q \VV^5 - 23 Q^2 \VV^5 + 20 \VV^6 \cr
& - 62 Q \VV^6 + 52 Q^2 \VV^6 - 12 Q^3 \VV^6 + 3 Q^4 \VV^6 - 35 \VV^7 + 140 Q \VV^7 \cr
&- 172 Q^2 \VV^7 + 67 Q^3 \VV^7 + 15 Q \VV^8 - 48 Q^2 \VV^8 + 42 Q^3 \VV^8 - 9 Q^4 \VV^8 \cr
& - 10 Q^2 \VV^9 + 22 Q^3 \VV^9 - 11 Q^4 \VV^9 - Q^5 \VV^9 - 8 Q^3 \VV^{10} + 12 Q^4 \VV^{10} \cr
& + 3 Q^4 \VV^{11} - 7 Q^5 \VV^{11} + 3 Q^5 \VV^{12} - 3 Q^6 \VV^{12}\big)\UU^{4}+\VV^{34}\big(3 - 3 Q + 5 \VV \cr
& - 7 Q \VV - 9 \VV^2 + 11 Q \VV^2 + 18 \VV^3 - 34 Q \VV^3 + 17 Q^2 \VV^3 - Q^3 \VV^3 + 5 \VV^4 \cr
& - 19 Q \VV^4 + 14 Q^2 \VV^4 - 15 \VV^5 + 38 Q \VV^5 - 25 Q^2 \VV^5 + 2 Q^3 \VV^5 + 21 \VV^6 \cr
& - 70 Q \VV^6 + 69 Q^2 \VV^6 - 21 Q^3 \VV^6 + 3 Q^4 \VV^6 - 6 Q \VV^7 + 17 Q^2 \VV^7 \cr
& - 13 Q^3 \VV^7 + 5 Q^2 \VV^8 - 4 Q^3 \VV^8 - Q^4 \VV^8 - 3 Q^3 \VV^9 + 6 Q^4 \VV^9 - 3 Q^5 \VV^9 \cr
& -  Q^5 \VV^{11} + Q^6 \VV^{12}\big)\UU^{5}+\VV^{40}\big(1 - \VV + 3 \VV^3 - 3 Q \VV^3 - 3 \VV^4 + 3 Q \VV^4 \cr
& + 5 \VV^5 - 5 Q \VV^5 + \VV^6 - 4 Q \VV^6 + 3 Q^2 \VV^6 - 6 \VV^7 + 9 Q \VV^7 - 3 Q^2 \VV^7 \cr
& + 7 \VV^8 - 14 Q \VV^8 + 7 Q^2 \VV^8 - Q \VV^9 + 2 Q^2 \VV^9 - Q^3 \VV^9 + Q^2 \VV^{10} \cr
& + Q^3 \VV^{10} - 2 Q^3 \VV^{11}\big)\UU^{6}-\VV^{55}(1-\VV)\UU^{7}\  .}}

\listrefs
\end